\def\zabs{$z_{\rm abs}$}

\def\sii{S~{\sc ii}}

\def\hi{H~{\sc i}}
\def\hii{H~{\sc ii}}

\def\feii{Fe~{\sc ii}}

\def\siii{Si~{\sc ii}}
\def\siiii{Si~{\sc iii}}

\def\alii{Al~{\sc ii}}

\def\oi{O~{\sc i}}

\def\ni{N~{\sc i}}

\documentclass[useAMS,usenatbib,epsfig]{mn2e}

\usepackage{rotating}
\usepackage{float}
\usepackage{natbib}
\usepackage{verbatim}
\usepackage{hyperref,graphicx}

\title[Evidence of Bimodality in the N/$\alpha$ Distribution]{The ESO UVES Advanced Data Products Quasar Sample - III. Evidence of Bimodality in the [N/$\alpha$] Distribution}
\author[T. Zafar et al.] {Tayyaba Zafar$^{1, 2}$\thanks{e-mail:tzafar@eso.org}, Miriam Centuri\'on$^3$, C\'eline P\'eroux$^2$, Paolo Molaro$^3$,  
\newauthor
Valentina D'Odorico$^3$, Giovanni Vladilo$^3$ \& Attila Popping$^4$\\
$^1$ European Southern Observatory, Karl-Schwarzschild-Strasse 2, 85748, Garching, Germany. \\
$^2$ Aix Marseille Universit\'e, CNRS, LAM (Laboratoire d'Astrophysique de Marseille) UMR 7326, 13388, Marseille, France.  \\
$^3$ INAF - Osservatorio Astronomico di Trieste, via G.B. Tiepolo 11, Trieste, Italy.\\
$^4$ International Centre for Radio Astronomy Research (ICRAR), The University of Western Australia, 35 Stirling Hwy, Crawley,\\ WA 6009, Australia.\\
}

\begin{document}


\pagerange{\pageref{firstpage}--\pageref{lastpage}} \pubyear{2013}

\maketitle

\label{firstpage}

\begin{abstract}
We report here a study of nitrogen and $\alpha$-capture element (O, S, and Si) abundances in 18 Damped Ly$\alpha$ Absorbers (DLAs) and sub-DLAs drawn from the ESO-UVES Advanced Data Products (EUADP) database. We report 9 new measurements, 5 upper and 4 lower limits of nitrogen that when compiled with available nitrogen measurements from the literature makes a sample of 108 systems. The extended sample presented here confirms the [N/$\alpha$] bimodal behaviour suggested in previous studies. Three-quarter of the systems show  $\langle$[N/$\alpha$]$\rangle=-0.85$ ($\pm$0.20\,dex) and one-quarter ratios are clustered at $\langle$[N/$\alpha$]$\rangle= -1.41$ ($\pm$0.14\,dex). The high [N/$\alpha$] plateau is consistent with the \hii\ regions of dwarf irregular and blue compact dwarf galaxies although extended to lower metallicities  and could be interpreted as the result of a primary nitrogen production by intermediate mass stars. The low [N/$\alpha$] values are the lowest ever observed in any astrophysical site. In spite of this fact, even lower values could be measured with the present instrumentation, but we do not find them below [N/$\alpha$] $\approx$ $-1.7$. This suggests the presence of a floor in [N/$\alpha$] abundances, which along with the lockstep increase of N and Si may indicate a primary nitrogen production from fast rotating, massive stars in relatively young or unevolved systems.
\end{abstract}

\section{Introduction}

The CNO cycle in the H-burning layer of stars is reasonably well understood as the main process for the nitrogen production. However, the characteristics of the stars which produce N, such as their masses, metallicities, and the yields are still not well established. Long-lived intermediate mass stars (IMS; 4 $\leq$ $M/M_\odot$ $\leq$ 8) are considered the main sites of N production. At high metallicity, the secondary nitrogen production is the dominant process in IMS, it occurs from the seed carbon and oxygen nuclei already present in the interstellar medium (ISM) out of which the star first condensed. In such case, the abundance of nitrogen is expected to be proportional to the square of the C, O abundances, showing a correlation in the classical N/O versus O/H diagram. At low metallicities primary N production in IMS occurs from the synthesis of carbon and oxygen freshly produced by the star in the He-burning shell. Thermal pulses occurring during the asymptotic giant branch (AGB) phase make possible the transport of the He-burning products to the H-burning shell producing primary N \citep{henry00,marigo01}. The signature of primary N production is expected to be an approximately constant N/O abundance with increasing O/H metallicity. In the case of massive stars, the N production is very uncertain due to the poor understanding of the transport mechanism of the seed C, O for the primary N synthesis \citep[see][]{woosley95,meynet02,chiappini05,maeder09}.


Studies of nitrogen abundances in systems spanning a wide range of metallicities (especially sites of low metallicities) may help in providing clues on the nucleosynthetic origin of nitrogen. Measurements of nitrogen have been performed in different astrophysical sites. In \hii\ regions of spiral and dwarf-irregular galaxies \citep{vanzee98,vanzee06}, metal-poor emission-line galaxies \citep{nava06,izotov06,perez09}, and \hii\ regions in blue compact dwarf (BCD) galaxies \citep{izotov99,izotov04}, (N/O) ratios show a primary plateau at low oxygen abundances and a secondary behaviour for [O/H] $>$ $-1$. Nitrogen abundances were also measured in low-metallicity galactic halo stars \citep{spite05} and \hii\ regions in Large Magellanic Cloud \citep{bekki10}. In particular, it is important to verify if there is any contribution from massive stars at low metallicities. In this framework, high redshift quasar absorbers, selected by their imprints on a background quasar spectrum, provide useful tools. The absorbers with large neutral hydrogen column densities, the Damped Ly$\alpha$ absorbers (DLAs; log N(\hi)$>$20.3) or sub-DLAs (19.0 $<$ log N(\hi) $<$ 20.3) are excellent tools to study the nitrogen production. Their typical metallicities of $-3.0$ $\leq$ $Z/Z_\odot$ $\leq$ $-0.5$ reach even lower values than the \hii\ regions of dwarf galaxies. However, studies of nitrogen abundances require an estimate of the nitrogen to oxygen ratio (N/O). Accurate oxygen and nitrogen measurements are difficult to derive in quasar absorbers because of the saturation of the lines available and because some of these are blended with the Ly$\alpha$ forest features. Nevertheless, a number of such measurements have been performed over the years \citep{pettini95,molaro96,pettini02,centurion03,molaro03,molaro04,molaro06,petitjean08, cooke11,dutta14}. Using other $\alpha$-elements (S or Si) together with oxygen in 32 DLAs (21 measurements), \citet{centurion03} claimed that in addition to the DLAs clustered around the primary plateau, there exist a population comprising 25\% of the DLAs which is lying below the primary plateau around [N/Si] $=-1.5$. They noticed that no system with low [N/Si] values are observed at [N/H] $\geq-2.8$ which indicates that this abundance dichotomy could be attributed to different nitrogen enrichment histories.

The aims of this paper are: $i)$ to directly test the existence of the [N/$\alpha$] bimodality of quasar DLAs, $ii)$ to test the presence of ``floor'' in the [N/$\alpha$] ratios, and $iii)$ to explore the nucleosynthetic origin of primary nitrogen. This is done by expanding the sample of DLAs/sub-DLAs with new measurements of nitrogen. The study, together with a compilation of measurements from the literature to date, composes the largest set of nitrogen abundances in DLAs and sub-DLAs. The paper is organised as follows. In \S 2 we present the data sample and in \S 3 we describe the N and $\alpha$ abundance measurements. The results are discussed in \S 4 followed by conclusions in \S 5.

\section{Data Sample}

\subsection{New Data}

The new \ni\ measurements are from a sample of 250 quasar spectra observed with the Ultraviolet Visual Echelle Spectrograph (UVES) \citet{dekker00} mounted on the European Southern Observatory's (ESO), Very Large Telescope (VLT). \citet{zafar13} have taken advantages of the ESO advanced data archive to build the so-called EUADP (ESO UVES Advanced Data Products) sample of high-resolution quasar spectra. The data were post-processed and analysed to look for the DLAs and sub-DLAs they contain. The search resulted in a sample of 197 DLAs and sub-DLAs, part of which was used to compute the statistical properties of quasar absorbers and the consequences on the cosmological evolution of neutral gas mass \citep{zafar13b}. The wavelength coverage of the EUADP spectra is such that it allows for the detection of \ni\ lines from 1.65 $<$ $z_{\rm abs}$ $<$ 4.74. We have used this dataset to search for \ni\ lines in 140 such DLAs and sub-DLAs. Among the data of high enough quality, 27 measurements and 10 limits of \ni\ and $\alpha$-capture element have already been reported in the literature. We derived 9 new measurements of \ni\ together with 5 upper and 3 lower limits. In addition, we have analysed the DLA towards Q\,0334$-$1612 which was part of an earlier UVES sub-DLAs study \citep{peroux05} and obtained \ni\ lower limit. In total, the new sample is therefore composed of 9 measurements, 5 upper and 4 lower limits.

\subsection{Literature Data}

To complement the data described above, we have made a complete reappraisal of the \ni\ measurements in DLAs and sub-DLAs published in the literature to date. We have compiled 53 (34 measurements, 19 limits) DLAs/sub-DLAs with nitrogen and $\alpha$-capture element measurements from the literature. These high-resolution data are mainly acquired with the ESO-UVES and Keck-High Resolution Echelle Spectrometer (HIRES) instruments. In particular, we have reanalysed some of the data with extreme abundance ratios for which better fitting of the \ni\ lines could be done. In most cases, our analysis confirmed the results published by original authors. In one case towards Q\,1108-077 (\zabs=3.608), we report a large difference. The \ni\ abundance of this DLA was first derived by \citet{petitjean08}. We have used the data available to us through the EUADP sample to reanalyse the fits. We measure log N(\ni) $<$13.22 from the 3$\sigma$ limit of the undetected $\lambda$ 1134.9\,\AA\ line by using the linear part of the curve of growth. \citet{petitjean08} reported log N(\ni) $<12.84$ but this upper limit seems to be too stringent given the S/N of about 10-15 from their spectrum (see their Fig. 9). Unfortunately, the authors did not report any further detail on their measurements. For this reason we choose to use our \ni\ limit in the subsequent analysis.

In addition, for a few of the \citet{cooke11} metal-poor systems, we interacted directly with the authors who performed new abundance measurements to optimise the fits of \ni, \oi\ and \siii\ in particular. For J\,1558-0031, \citet{omeara06} published [N/Si] $=-0.10$ and [N/S] $=-0.32$ which is probably related to an issue with the estimate of N(\siii). R. Cooke measures [N/Si] $= -0.36$ and  [N/S] $= -0.29$ from the same data (private communication). We use the latter values in our analysis. R. Cooke also provided a revised measure of \oi\ in the system towards Q\,0913+072. Finally, new estimates of \ni\ and \oi\ in the absorber towards J\,1419+0829 are included in our analysis. Note that J\,0035-0918 is a carbon-enhanced metal-poor DLA. In this work, \citet{cooke14} abundances for this system are used \citep[see also][]{dutta14}.

\section{Column Density Determination}

\subsection{Method}

\subsubsection{Fitting Procedure}

The measurements of \ni\ are typically based on the two triplets around $\lambda1134$ and $\lambda1200$\,\AA. A total of six transitions i.e., $i)$ \ni\ ($\lambda\lambda\lambda$ 1134.1, 1134.4, 1134.9) and $ii)$ \ni\ ($\lambda\lambda\lambda$ 1199, 1200.2, 1200.7) are often detected in quasar absorbers. The number of lines from the two triplets which are fitted are maximised in our analysis. The \ni\ column densities (and other ions - \sii, \siii, etc - in some cases) have been derived using the $\chi^2$ minimisation routine \texttt{FITLYMAN} package within the \texttt{MIDAS} environment \citep{fontana95}. Laboratory wavelengths and oscillator strengths were taken from \citet{morton03}. The global fit returns the best fit parameters for the central wavelength, column density and Doppler turbulent broadening, as well as errors on each quantity, including 1$\sigma$ errors on the column densities. 

Given that all six \ni\ transitions are located in the Ly$\alpha$ forest, great care was taken to assess the contamination of the lines by interlopers. This possible blending would tend to overestimate the column density of \ni. A comparison with radial velocity profiles of low ionisation species (\siii\, \sii\, \feii\ and \alii) has been made to ascertain the transitions free from contamination. These low ionisation species were chosen in order to obtain the $\alpha$-element abundance from different ions or as reference absorption lines in case they lie outside the Ly$\alpha$ forest and thus free from possible blending. Upper limits were derived in cases with evidences for contamination of the \ni\ transitions by Ly$\alpha$ forest interlopers. 

\subsubsection{Effects of Photoionisation}\label{sec:phot}

While the DLAs are largely neutral in hydrogen, it is expected that sub-DLAs contain some fractions of ionised gas, especially the systems with the lowest N(\hi) column densities. Ionisation correction of sub-DLAs may be a concern for deriving column densities. However, the neutral species are linked through the efficient charge exchange reactions and make the ionisation corrections negligible for log N(\hi) $>19.5$ \citep{viegas95}. Nevertheless, in case of penetration of enough hard photons, $\alpha$-elements and nitrogen could be over ionised compared to \hi. For most DLAs/sub-DLAs, the predominant source of hard photons is the background ionisation field. \citet{meiring09} used photoionisation models based on such field over the sub-DLA N(\hi) range and show that the ionisation corrections to the column densities are in general small. They find that a strong ionising field with ionising parameter, log $U<-2.5$, could increase the measured abundances by at most $-0.2$ dex. From an investigation of 20 DLAs, \citet{prochaska02} reported that the ionisation correction for [N/$\alpha$] is at most +0.1\,dex. In our analysis, only three new measures of \ni\ in sub-DLAs are reported (for \zabs=2.010 towards B\,0122-005, for \zabs=3.385 towards J\,0332-4455, and for \zabs=2.113 towards B\,1220-1800) and most of the sub-DLAs have log N(\hi) $>$20.0. Given that the corrections are as large as the uncertainties, we did not apply any modifications to the reported abundances.

\subsection{Notes on Individual Systems}

In the following, each system for which \ni\ column density estimate is made for the first time in this study is briefly described. The details of the UVES spectra used in this work and \hi\ column density references for each system are provided in \citet{zafar13,zafar13b}. For a given absorption system, the abundance of each element is calculated by adding the column densities of all components.

\begin{figure}
\begin{center}
{\includegraphics[width=\columnwidth,clip=]{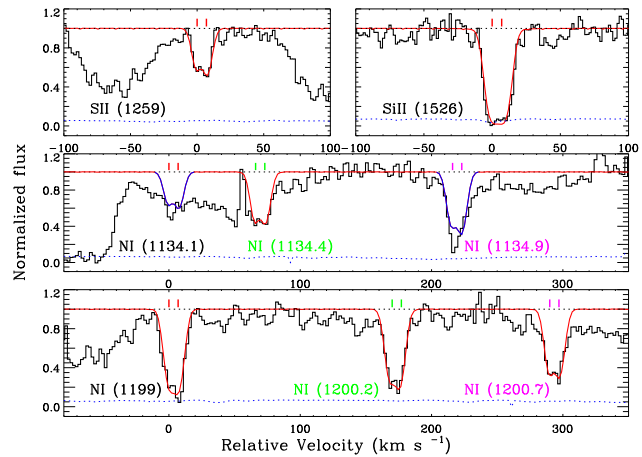}}
\caption{Voigt-profile fits (red overlay) of \ni\ and \sii\ for the DLA system at $z_{abs}=2.3697$ (zero velocity) in J\,0105-1846. Normalised quasar continuum and error spectrum are shown in black and blue dotted lines, respectively. The blue overlays represent the profile of the transitions that were not fit and obtained from the fit of the other transitions in the system. The red tick marks above the normalised continuum indicate the locations of the velocity components. The green and magenta tick marks and labels correspond to the second and third transitions in each multiplet.}
\label{fig:0105}
\end{center}
\end{figure}

\underline{J\,0105-1846, \zabs=2.3697:} the two-component structure of the system is derived from the fit to the \sii\ $\lambda$ 1259 absorption line. The nitrogen abundance is derived from unsaturated, non-blended \ni\ $\lambda\lambda\lambda\lambda$ 1134.4, 1199, 1200.2, and 1200.7 lines (see Fig. \ref{fig:0105}). The properties of the velocity components are described in Table \ref{tab:0105}. Note that \citet{srianand05} derived log N(\sii)$=14.30\pm0.04$, consistent to our results within the errors. The \siii\ lines are either saturated or blended. We derive a lower limit from \siii\ $\lambda$ 1526 transition: log N(\siii)$>14.86$, by using the redshift and $b$-value of the two components obtained from the \sii\ transitions.

\begin{table}
\caption{Component structure of the $z_{\rm abs}=2.3697$ DLA towards J\,0105-1846.}
\centering
\label{tab:0105}
\begin{tabular}{l c c c c}
\hline\hline                       
Comp. & $z_{\rm abs}$ & $b$ & Ion & log $N$ \\
	 & & km s$^{-1}$ & & cm$^{-2}$ \\ 
\hline
1 & 2.3696 & $1.8\pm0.6$ & \sii\ & $14.04\pm0.04$ \\
    &              &                         & \siii\ & $>14.74$ \\
    &              &                         & \ni\ & $14.14\pm0.03$ \\
2 & 2.3697 & $2.5\pm0.8$ & \sii\ & $14.10\pm0.03$ \\
    &              &                         & \siii\ & $>14.24$ \\
    &              &                         & \ni\ & $14.08\pm0.02$ \\
\hline
\end{tabular}
\end{table}

\begin{figure}
\begin{center}
{\includegraphics[width=\columnwidth,clip=]{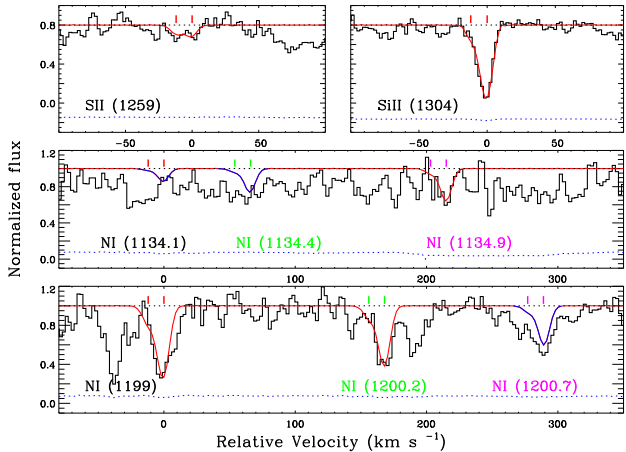}}
\caption{Voigt-profile fits (red overlay) of \ni, \sii, and \siii\ for the sub-DLA at $z_{abs}=2.0095$ (zero velocity) in B\,0122-005.}
\label{fig:0122}
\end{center}
\end{figure}

\underline{B\,0122-005, \zabs=2.0095:} the line profiles of this sub-DLA are characterised by a simple velocity structure composed of two components. The structure of the system is derived from the fit to the \sii\ $\lambda$ 1259 and \siii\ $\lambda$ 1304 absorption lines. The \ni\ absorption is seen in the $\lambda\lambda\lambda$ 1134.9, 1199, and 1200.2 lines (see Fig. \ref{fig:0122}). The properties of the velocity components are described in Table \ref{tab:0122}. Note that \citet{ellison09} derived log N(\siii)$=13.67\pm0.05$, which is in good agreement with our result.

\begin{table}
\caption{Component structure of the $z_{\rm abs}=2.0095$ sub-DLA towards B\,0122-005.}
\centering
\label{tab:0122}
\begin{tabular}{l c c c c}
\hline\hline                       
Comp. & $z_{\rm abs}$ & $b$ & Ion & log $N$ \\
	 & & km s$^{-1}$ & & cm$^{-2}$ \\ 
\hline
1 & 2.0094 & $4.7\pm1.2$ & \sii\ & $13.30\pm0.06$ \\
    &              &                         & \siii\ & $12.77\pm0.05$ \\
    &              &                         & \ni\ & $12.79\pm0.03$ \\
2 & 2.0095 & $3.8\pm1.0$ & \sii\ & $13.34\pm0.06$ \\
    &              &                         & \siii\ & $13.63\pm0.06$ \\
    &              &                         & \ni\ & $13.46\pm0.04$ \\
\hline
\end{tabular}
\end{table}

\begin{figure}
\begin{center}
{\includegraphics[width=\columnwidth,clip=]{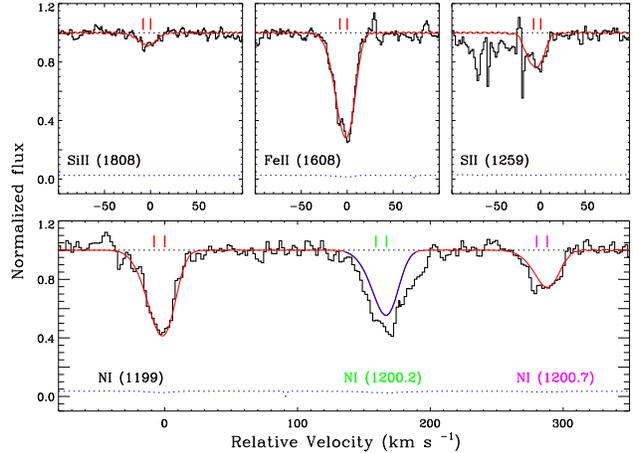}}
\caption{Voigt-profile fits (red overlay) of \ni, \siii, and \feii\ for the sub-DLA at $z_{abs}=2.4112$ (zero velocity) in J\,0332-4455.}
\label{fig:0332}
\end{center}
\end{figure}

\underline{J\,0332-4455, \zabs=2.4112:} the line profiles of this sub-DLA are characterised by a simple velocity structure composed of two components. The structure of the system is derived from the fit of \sii\ $\lambda$ 1259, \feii\ $\lambda\lambda$ 1608 and 1611, and \siii\ $\lambda$ 1808 absorption lines. The \ni\ absorption is seen in two lines $\lambda\lambda$ 1199 and 1200.7 of the higher order triplet (see Fig. \ref{fig:0332}). The properties of the velocity components are described in Table \ref{tab:0332}. 

\begin{table}
\caption{Component structure of the $z_{\rm abs}=2.4112$ sub-DLA towards J\,0332-4455.}
\centering
\label{tab:0332}
\begin{tabular}{l c c c c}
\hline\hline                       
Comp. & $z_{\rm abs}$ & $b$ & Ion & log $N$ \\
	 & & km s$^{-1}$ & & cm$^{-2}$ \\
\hline
1 & 2.4111 & $13.9\pm3.5$ & \sii\ & $14.06\pm0.03$ \\
    &              &                         & \siii\ & $14.05\pm0.03$ \\
    &              &                         & \feii\ & $13.56\pm0.02$ \\
    &              &                         & \ni\ & $13.29\pm0.04$ \\
2 & 2.4112 & $10.1\pm1.3$ & \sii\ & $13.38\pm0.02$ \\
    &              &                           & \siii\ & $14.00\pm0.03$ \\
    &              &                         & \feii\ & $13.89\pm0.02$ \\
    &              &                         & \ni\ & $13.46\pm0.03$ \\
\hline
\end{tabular}
\end{table}

\begin{figure}
\begin{center}
{\includegraphics[width=\columnwidth,clip=]{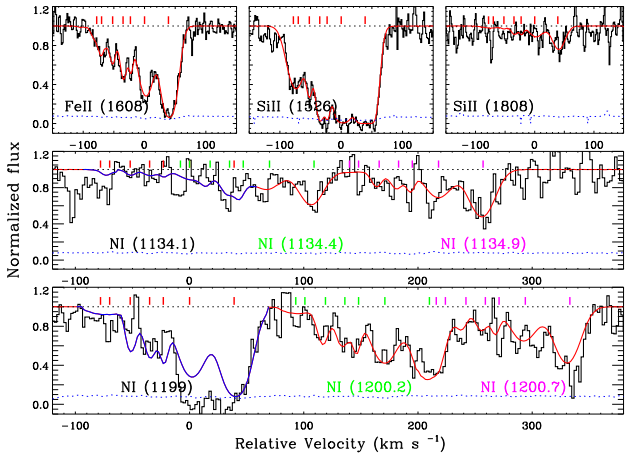}}
\caption{Voigt-profile fits (red overlay) of \ni, \feii, and \siii\ for the DLA system at $z_{abs}=2.3016$ (zero velocity) in B 0432-440.}
\label{fig:0432}
\end{center}
\end{figure}

\underline{B\,0432-440, \zabs=2.3016:} the velocity profiles of the metal lines of this DLA are extremely complex and spread over 190 km s$^{-1}$. We derive seven main components by fitting simultaneously the velocity profiles of \feii\ $\lambda$ 1608, \siii\ $\lambda\lambda$ 1526 and 1808 absorption lines. Because of the complex structure, \ni\ absorption lines are blended with each other. However, all seven \ni\ velocity components are clearly seen in $\lambda\lambda\lambda$ 1134.9, 1200.2, and 1200.7 transitions (see Fig. \ref{fig:0432}). The redshift, $b$-value, and column density of each component are given in Table \ref{tab:0432}. For \siii\ we obtain log N(\siii)$=$$15.11\pm0.13$ from \siii 1526 and 1808 lines, which is consistent with the column densities obtained by \citet{akerman05} (log N(\siii)$=$$15.20$) and \citet{noterdaeme08} (log N(\siii)$=$$15.27\pm0.07$).

\begin{table}
\caption{Component structure of the $z_{\rm abs}=2.3016$ DLA towards B\,0432-440.}
\centering
\label{tab:0432}
\begin{tabular}{l c c c c}
\hline\hline                       
Comp. & $z_{\rm abs}$ & $b$ & Ion & log $N$ \\
	 & & km s$^{-1}$ & & cm$^{-2}$ \\ 
\hline
1 & 2.3007 & $11.5\pm2.5$ & \feii\ & $13.35\pm0.02$ \\
    &              &                         & \siii\ & $13.01\pm0.02$ \\
    &              &                         & \ni\ & $12.79\pm0.03$ \\
2 & 2.3008 & $9.2\pm1.1$ & \feii\ & $13.45\pm0.02$ \\
    &              &                         & \siii\ & $13.84\pm0.03$ \\
    &              &                         & \ni\ & $12.97\pm0.02$ \\
3 & 2.3010 & $5.0\pm0.9$ & \feii\ & $13.25\pm0.02$ \\
    &              &                         & \siii\ & $13.40\pm0.07$ \\
    &              &                         & \ni\ & $13.25\pm0.03$ \\
4 & 2.3012 & $7.0\pm1.1$ & \feii\ & $13.65\pm0.01$ \\
    &              &                         & \siii\ & $14.08\pm0.04$ \\
    &              &                         & \ni\ & $13.39\pm0.03$ \\
5 & 2.3013 & $3.5\pm0.7$ & \feii\ & $13.25\pm0.02$ \\
    &              &                         & \siii\ & $13.81\pm0.09$ \\
    &              &                         & \ni\ & $13.19\pm0.04$ \\
6 & 2.3016 & $15.2\pm0.9$ & \feii\ & $14.15\pm0.01$ \\
    &              &                         & \siii\ & $14.51\pm0.03$ \\
    &              &                         & \ni\ & $13.95\pm0.01$ \\
7 & 2.3020 & $13.8\pm1.2$ & \feii\ & $14.45\pm0.01$ \\
    &              &                         & \siii\ & $14.83\pm0.03$ \\
    &              &                         & \ni\ & $14.17\pm0.01$ \\
\hline
\end{tabular}
\end{table}

\begin{figure}
\begin{center}
{\includegraphics[width=\columnwidth,clip=]{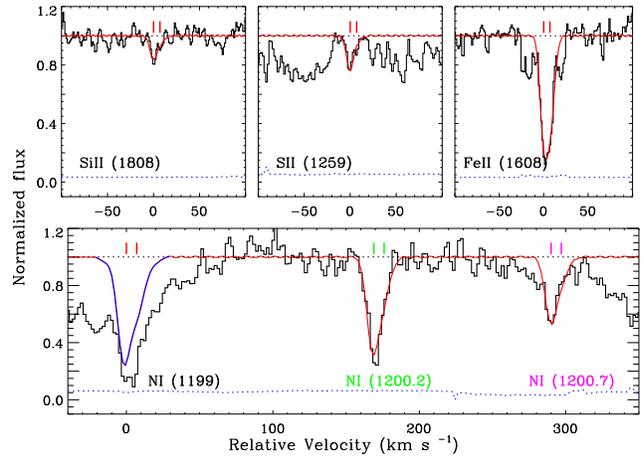}}
\caption{Voigt-profile fits (red overlay) of \ni, \siii, \sii, and \feii\ for the DLA system at $z_{abs}=2.6077$ (zero velocity) in J\,1155+0530.}
\label{fig:1155}
\end{center}
\end{figure}

\underline{J\,1155+0530, \zabs=2.6077:} the \siii\ $\lambda$ 1808, \sii\ $\lambda$ 1259, and \feii\ $\lambda$ 1608 metal lines in our data are well fitted by a simple two-component model (see Fig. \ref{fig:1155}). The Nitrogen column density is obtained from the analysis of the unsaturated, non-blended $\lambda\lambda$ 1200.2 and 1200.7 transitions. The column density, redshift, and $b$-value of each component is given in Table \ref{tab:1155}.
\begin{table}
\caption{Component structure of the $z_{\rm abs}=2.6077$ DLA towards J\,1155+0530.}
\centering
\label{tab:1155}
\begin{tabular}{l c c c c}
\hline\hline                       
Comp. & $z_{\rm abs}$ & $b$ & Ion & log $N$ \\
	 & & km s$^{-1}$ & & cm$^{-2}$ \\ 
\hline
1 & 2.6077 & $3.0\pm0.6$ & \sii\ & $13.64\pm0.02$ \\   
    &              &                         & \siii\ & $14.11\pm0.04$ \\
    &              &                         & \feii\ & $13.69\pm0.02$ \\
    &              &                         & \ni\ & $13.60\pm0.04$ \\
2 & 2.6077 & $5.7\pm0.9$ & \sii\ & $13.06\pm0.02$ \\
    &              &                         & \siii\ & $13.88\pm0.03$ \\
    &              &                         & \feii\ & $13.85\pm0.02$ \\
    &              &                         & \ni\ & $13.24\pm0.05$ \\
\hline
\end{tabular}
\end{table}

\begin{figure}
\begin{center}
{\includegraphics[width=\columnwidth,clip=]{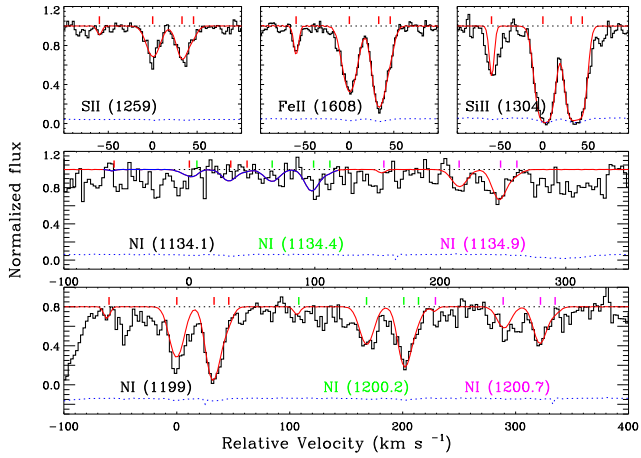}}
\caption{Voigt-profile fits (red overlay) of \ni, \sii, and \feii\ for the sub-DLA at $z_{abs}=2.1125$ (zero velocity) in B\,1220-1800.}
\label{fig:1220}
\end{center}
\end{figure}

\underline{B\,1220-1800, \zabs=2.1125:} the absorption line profiles of this sub-DLA are characterised by four components spread over 110 km s$^{-1}$. The structure of the system is derived from the fit to the \sii\ $\lambda$ 1259 and \feii\ $\lambda$ 1608 transitions. The nitrogen abundance is derived from unsaturated, non-blended \ni\ $\lambda\lambda\lambda\lambda$ 1134.9, 1199, 1200.2, and 1200.7 absorption lines (see Fig. \ref{fig:1220}). The properties of each velocity components are given in Table \ref{tab:1220}. \ni\ have been fitted without tiding the $b$ and $z$ parameters to the other element transitions in a higher state of ionization (i.e., \sii, \siii, and \feii). We obtained the same $z$ and $b$ values consistent within uncertainties. Previously, \citet{noterdaeme08} derived log N(\sii)$=14.39\pm0.03$ which is in good agreement with our result. The \siii\ lines are either saturated or have low signal-to-noise ratio. We derive a lower limit from \siii\ $\lambda$ 1304 transition: log N(\siii)$>14.75$, by using the redshift and $b$-value of the four components obtained from the \sii\ and \feii\ transitions. 
\begin{table}
\caption{Component structure of the $z_{\rm abs}=2.1125$ sub-DLA towards B\,1220-1800.}
\centering
\label{tab:1220}
\begin{tabular}{l c c c c}
\hline\hline                       
Comp. & $z_{\rm abs}$ & $b$ & Ion & log $N$ \\
	 & & km s$^{-1}$ & & cm$^{-2}$ \\ 
\hline
1 & 2.1119 & $1.4\pm0.2$ & \sii\ & $13.03\pm0.02$ \\
    &              &                         & \siii\ & $13.51\pm0.02$ \\
    &              &                         & \feii\ & $13.07\pm0.02$ \\
    &              &  $1.4\pm0.2$  & \ni\ & $12.37\pm0.03$ \\
2 & 2.1125 & $9.6\pm1.0$ & \sii\ & $14.03\pm0.02$ \\
    &              &                         & \siii\ & $>14.45$ \\
    &              &                         & \feii\ & $13.92\pm0.01$ \\
    &              & $7.7\pm1.2$ & \ni\ & $13.48\pm0.01$ \\
3 & 2.1128 & $8.0\pm0.5$ & \sii\ & $13.98\pm0.01$ \\
    &              &                         & \siii\ & $>14.21$ \\
    &              &                         & \feii\ & $14.05\pm0.01$ \\
    &              & $6.7\pm0.8$ & \ni\ & $13.68\pm0.02$ \\
4 & 2.1130 & $6.2\pm0.4$ & \sii\ & $13.37\pm0.03$ \\
    &              &                         & \siii\ & $13.95\pm0.02$ \\
    &              &                         & \feii\ & $13.20\pm0.02$ \\
    &              &  $6.0\pm0.5$ & \ni\ & $12.83\pm0.05$ \\
\hline
\end{tabular}
\end{table}

\begin{figure}
\begin{center}
{\includegraphics[width=\columnwidth,clip=]{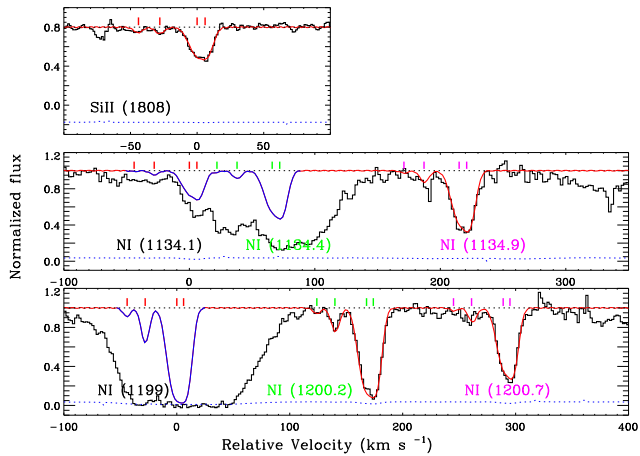}}
\caption{Voigt-profile fits (red overlay) of \ni\ and \siii\ for the DLA system at $z_{abs}=2.8523$ (zero velocity) in LBQS\,2138-4427.}
\label{fig:2138}
\end{center}
\end{figure}

\underline{LBQS\,2138-4427, \zabs=2.8523:} the absorption line profiles of this sub-DLA are characterised by a clear four components spread over 70 km s$^{-1}$. This profile is derived from the fit of \siii\ $\lambda$ 1808 transition. The Nitrogen abundance is derived from unsaturated, non-blended \ni\ $\lambda\lambda\lambda$ 1134.9, 1200.2, and 1200.7 transitions (see Fig. \ref{fig:2138}). The column density, redshift, and $b$-parameter of each velocity component is given in Table \ref{tab:2138}. Note that \citet{srianand05} derived log N(\siii)$=14.86\pm0.02$ and log N(\sii)$=14.50\pm0.02$, where former is in good agreement with our result. 

\begin{table}
\caption{Component structure of the $z_{\rm abs}=2.8523$ DLA towards LBQS\,2138-4427.}
\centering
\label{tab:2138}
\begin{tabular}{l c c c c}
\hline\hline                       
Comp. & $z_{\rm abs}$ & $b$ & Ion & log $N$ \\
	 & & km s$^{-1}$ & & cm$^{-2}$ \\ 
\hline
1 & 2.8517 & $2.0\pm0.4$ & \siii\ & $13.68\pm0.02$ \\
    &              &                         & \ni\ & $12.27\pm0.03$ \\
2 & 2.8519 & $2.8\pm0.7$ & \siii\ & $13.76\pm0.02$ \\
    &              &                         & \ni\ & $13.02\pm0.02$ \\
3 & 2.8523 & $7.2\pm0.8$ & \siii\ & $14.64\pm0.01$ \\
    &              &                         & \ni\ & $14.03\pm0.01$ \\
4 & 2.8524 & $3.8\pm0.6$ & \siii\ & $14.32\pm0.01$ \\
    &              &                         & \ni\ & $13.82\pm0.02$ \\
\hline
\end{tabular}
\end{table}

\begin{figure}
\begin{center}
{\includegraphics[width=\columnwidth,clip=]{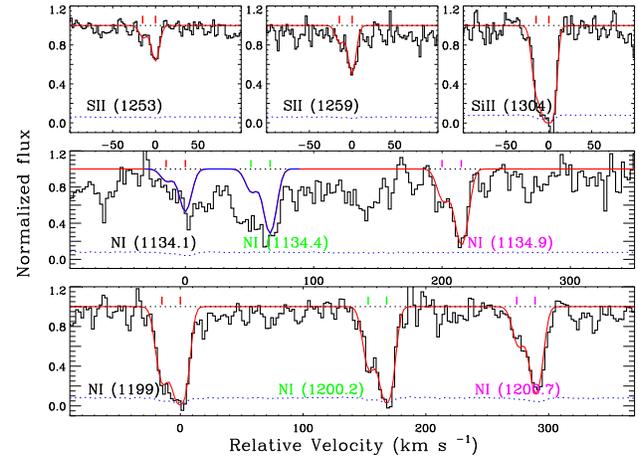}}
\caption{Voigt-profile fits (red overlay) of \ni\ and \sii\ for the DLA system at $z_{abs}=2.1539$ (zero velocity) in B\,2222-396.}
\label{fig:2222}
\end{center}
\end{figure}

\underline{B\,2222-396, \zabs=2.1539:} the line profiles of this DLA are characterised by a simple velocity structure composed of two components. The structure of the system is derived from the fit of \sii\ $\lambda\lambda$ 1253 and 1259 absorption lines. The \ni\ absorption is seen in four transitions $\lambda\lambda\lambda\lambda$ 1134.9, 1199, 1200.2 and 1200.7 of the two triplets (see Fig. \ref{fig:2222}). The properties of the velocity components are described in Table \ref{tab:2222}. Note that \citet{noterdaeme08} obtained log N(\sii)$=14.08\pm0.07$ which is consistent with our result. The \siii\ lines are saturated and we derive a lower limit from \siii\ $\lambda$ 1304 transition: log N(\siiii)$>15.04$, by using the redshift and $b$-value of the two components obtained from the \sii\ transitions.

\begin{table}
\caption{Component structure of the $z_{\rm abs}=2.1539$ DLA towards B\,2222-396.}
\centering
\label{tab:2222}
\begin{tabular}{l c c c c}
\hline\hline                       
Comp. & $z_{\rm abs}$ & $b$ & Ion & log $N$ \\
	 & & km s$^{-1}$ & & cm$^{-2}$ \\ 
\hline
1 & 2.1537 & $5.4\pm0.6$ & \sii\ & $13.63\pm0.04$ \\
    &              &                         & \siii\ & $>13.69$ \\
    &              &                         & \ni\ & $13.63\pm0.03$ \\
2 & 2.1539 & $5.1\pm0.4$ & \sii\ & $14.13\pm0.04$ \\
    &              &                         & \siii\ & $>15.02$ \\
    &              &                         & \ni\ & $14.25\pm0.04$ \\
\hline
\end{tabular}
\end{table}

\begin{figure}
\begin{center}
{\includegraphics[width=\columnwidth,clip=]{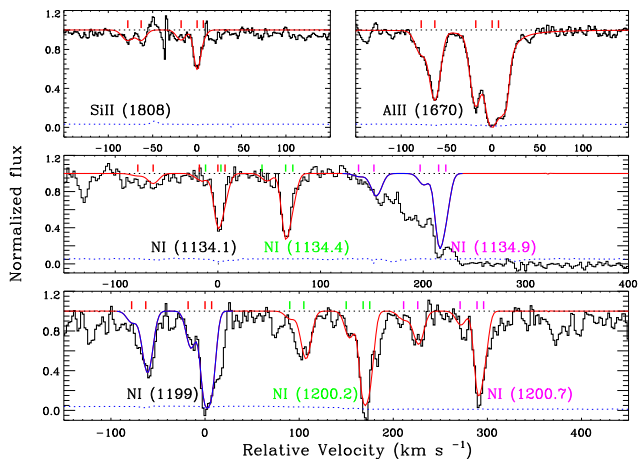}}
\caption{Voigt-profile fits (red overlay) of \ni, \siii, and \alii\ for the DLA system at $z_{abs}=2.6147$ (zero velocity) in LBQS\,2348-0180.}
\label{fig:2348}
\end{center}
\end{figure}

\underline{B\,2348-0180, \zabs=2.6147:} the line profiles of this DLA are characterised by five main components spread over 130 km s$^{-1}$. The component structure of the system is derived from the fit of \siii\ $\lambda$ 1808 and \alii\ $\lambda$ 1670 transitions. The nitrogen abundance is derived from \ni\ $\lambda\lambda\lambda\lambda$ 1134.1, 1134.4, 1200.2, and 1200.7 transitions (see Fig. \ref{fig:2348}). The column density, redshift, and $b$-parameter of each component is given in Table \ref{tab:2348}. Previously \citet{prochaska01} derived log N(\alii)$>13.14$ and log N(\siiii)$=14.89\pm0.07$ from Keck/HIRES data, where the latter is consistent with our result. The resulting total column densities for each systems are provide in Table \ref{column_density}.

\begin{table}
\caption{Component structure of the $z_{\rm abs}=2.6147$ DLA towards B\,2348-0180.}
\centering
\label{tab:2348}
\begin{tabular}{l c c c c}
\hline\hline                       
Comp. & $z_{\rm abs}$ & $b$ & Ion & log $N$ \\
	 & & km s$^{-1}$ & & cm$^{-2}$ \\ 
\hline
1 & 2.6138 & $7.3\pm1.8$ & \siii\ & $14.21\pm0.03$ \\
   &               &                        & \alii\ & $11.70\pm0.03$ \\
   &               &                         & \ni\ & $12.99\pm0.03$ \\
2 & 2.6140 & $6.1\pm0.5$ & \siii\ & $14.15\pm0.03$ \\
    &              &                         & \alii\ & $12.33\pm0.02$ \\
    &              &                         & \ni\ & $13.58\pm0.02$ \\
3 & 2.6145 & $6.0\pm0.7$ & \siii\ & $14.13\pm0.02$ \\
   &               &                         & \alii\ & $12.37\pm0.02$ \\
   &               &                         & \ni\ & $13.15\pm0.02$ \\
4 & 2.6147 & $2.9\pm0.3$ & \siii\ & $14.67\pm0.02$ \\
   &               &                         & \alii\ & $14.46\pm0.03$ \\
   &               &                         & \ni\ & $14.23\pm0.02$ \\
5 & 2.6148 & $4.0\pm0.4$ & \siii\ & $13.90\pm0.03$ \\
   &               &                        & \alii\ & $15.03\pm0.03$ \\
   &               &                         & \ni\ & $13.73\pm0.02$ \\
\hline
\end{tabular}
\end{table}

\section{Results and Discussion}
\subsection{The Origin of Nitrogen}
It is believed that the main pathway for the production of N in stars is a six-step process in the CN branch of the CNO cycle which happens in the stellar H-burning layer, with the result that $^{14}$N is synthesised from $^{12}$C and $^{16}$O. The nitrogen production in stars of different masses and metallicities is not well established. Nitrogen has two production pathways either `primary' or `secondary',  depending on whether seed carbon and oxygen are produced by the star itself (primary) or they were already present in the gas out of which the star was formed (secondary).

Secondary production is the standard process for Nitrogen production in the H-burning layers of stars of all masses (mainly from IMS) as the seeds $^{12}$C and $^{16}$O were already present in these layers. The secondary production dominates at high metallicities ([O/H] $>$ $-1$) and shows a correlation between (N/O) and (O/H).

The N-production duality is revealed when the measurements of N in \hii\ regions of spiral \citep{vanzee98}, BCD \citep{izotov99,izotov04}, and emission-line galaxies \citep{nava06,izotov06,perez09} are compared. In low metallicity environments N goes in lockstep with O, so that (N/O) remains approximately constant. This primary N is thought to be synthesised by long-lived IMS (4 $\leq$ $M/M_\odot$ $\leq$ 8) during the AGB phase, suffering dredged-up episodes, and hot-bottom burning processes \citep{henry00,marigo01,chiappini03}. The carbon then penetrates into the H-burning upper shell where it is transformed into nitrogen by the CNO cycle. There exist large uncertainties about the amount of primary nitrogen produced through the hot-bottom burning process. On the other hand, there is no clear mechanism that may produce primary N in massive stars. Only at higher metallicities, massive stars (M $\geq$20M$_\odot$) are foreseen to produce N \citep{woosley95}. At very low metallicities, only fast rotating massive stars (9 $\leq$ $M/M_\odot$ $\leq$ 20) can provide mixing of CNO into the He-burning shell \citep{meynet02,chiappini06}. \citet{chieffi02} have achieved this by introducing winds and high cross-sections to their models in the reaction leading to O. These yields of primary N production are still uncertain, since they are computed assuming a very high rotation velocity for the metal-poor stars. The discovery of metal-poor halo stars with high (N/O) ratios seems to confirm the primary N production in massive stars \citep{israelian04,spite05} with a yield that depends on stellar mass and metallicity \citep{chiappini06}.

The theoretical prescriptions regarding the primary N production are uncertain and model dependent, therefore the relative amount of primary N produced in massive stars \citep{chiappini05,maeder09} and IMS is not well established. Moreover, in the framework of chemical evolution models, the production of nitrogen depends on star-formation efficiencies and the stellar lifetimes. This further increases the dispersion in the prediction of these models \citep{molla06}.

\begin{table*}
\caption{DLAs/sub-DLAs metal column densities derived in this work.}
\setlength{\tabcolsep}{6.0pt}
\centering
\label{column_density}
\begin{tabular}{l c c c c c c c }
\hline\hline                       
QSO & $z_{abs}$ &  N(\hi) & N(\ni) & N(\sii) & N(\siii) & N(\feii) & N(\alii) \\
	 & & cm$^{-2}$ 	& cm$^{-2}$ & cm$^{-2}$ & cm$^{-2}$ & cm$^{-2}$ & cm$^{-2}$  \\ 
\hline
0105-1846 & 2.369 & $21.00\pm0.08$ & $14.26\pm0.04$ & $14.37\pm0.05$ & $>14.86$ & $\cdots$ & $\cdots$ \\
0122-005   & 2.010 & $20.04\pm0.07$ & $13.54\pm0.05$ & $13.62\pm0.04$ & $13.69\pm0.05$ & $\cdots$ & $\cdots$\\
0332-4455 & 2.411 & $20.15\pm0.07$ & $13.68\pm0.05$ & $14.14\pm0.04$ & $14.33\pm0.04$ & $14.06\pm0.03$ & $\cdots$ \\
0432-440   & 2.302 & $20.95\pm0.10$ & $14.49\pm0.07$ & $\cdots$ & $15.11\pm0.13$ & $14.74\pm0.04$ & $\cdots$ \\
1155+0530 & 2.608 & $20.37\pm0.11$ & $13.76\pm0.06$ & $13.74\pm0.03$	& $14.31\pm0.05$ & $14.08\pm0.03$ & $\cdots$ \\
1220-1800 & 2.113 & $20.12\pm0.07$ & $13.93\pm0.06$ & $14.37\pm0.04$ & $>14.75$ & $14.35\pm0.03$ & $\cdots$ \\
2138-4427 & 2.852 & $20.98\pm0.05$ & $14.27\pm0.04$ & $\cdots$ & $14.86\pm0.02$ & $\cdots$ & $\cdots$ \\
2222-396   & 2.154 & $20.85\pm0.10$ & $14.34\pm0.05$ & $14.25\pm0.06$ & $>15.04$ & $\cdots$ & $\cdots$ \\
2348-0180 & 2.615 & $21.30\pm0.08$ & $14.46\pm0.05$ & $\cdots$ & $15.00\pm0.06$ & $\cdots$ & $15.14\pm0.06$ \\ 
\hline
\end{tabular}
\end{table*}

High-redshift DLAs/sub-DLAs can add valuable information on the overall picture of nitrogen production, since they span a broad range of metallicities, $-3.0$ $\leq$ $Z/Z_\odot$ $\leq$ $-0.5$, reaching metallicities $-1.5$\,dex lower than those of \hii\ regions. Starting from the first studies \citep{pettini95,molaro96}, the number of DLAs/sub-DLAs with reliable estimates of N has increased over the years. We here use a compilation of literature data together with new N estimates in order to explore the rare low-metallicity regime and study the early enrichment of nitrogen.

\begin{table}
\caption{Solar photospheric abundances used in this study (from \citealt{asplund09}).}
\centering
\label{solar}
\begin{tabular}{l c }
\hline\hline                       
Element & Photospheric abundance \\
\hline
H & 12.00\\
N & $7.83\pm0.05$ \\
O & $8.69\pm0.05$ \\
Si & $7.51\pm0.03$ \\
S & $7.12\pm0.03$ \\
Fe & $7.50\pm0.04$ \\
Zn & $4.56\pm0.05$ \\ 
\hline
\end{tabular}
\end{table}

\subsection{Abundance Ratios}
Our compilation of high-resolution literature data and new measurements reported here results in a sample of 70 \ni\ measurements, plus 38 upper and lower limits, making 108 in total. Using the measured column densities, we derive abundance ratios relative to solar. We adopted the solar photospheric abundances from \citealt{asplund09} (see Table \ref{solar}). We note that recent new measurements of \oi\ lines in stars by \cite{caffau13} might affect the reported oxygen solar abundance. The [N/$\alpha$] abundance ratio is simply the ratio of the nitrogen to the $\alpha$-capture elements. Referring to the solar abundances, [N/$\alpha$]=(N/$\alpha$)$-$(N/$\alpha)_\odot$, where (N/$\alpha$)=log $N$(\ni)$-$log $N(\alpha)$. The properties of each system are provided in Table \ref{N:abund}, including \zabs, \hi, \ni, $\alpha$-element column densities and abundance ratios [$\alpha$/H] and [N/$\alpha$]. While estimating abundance ratio limits, whenever the column density of one element is a limit, the other element is also treated as a limit at the same C.L. (i.e. 3$\sigma$) to be conservative with upper/lower bounds.

The choice of $\alpha$-element varies according to the spectroscopic data available for each absorber. Oxygen is the most suitable reference for the N abundances, since like nitrogen, oxygen is undepleted onto dust grains. Another important advantage of using oxygen is that the [N/O] and [O/H] ratios are derived from the observed \ni/\oi\ and \oi/\hi\ ratios. Therefore, these ratios are not affected by ionisation due to the presence of an intervening \hii\ region. However, the N(\oi) obtained from the strongly saturated \oi\ $\lambda$ 1302 \AA\, transition, results mostly in lower limits. High-resolution and high signal-to-noise spectra covering the blue part of wavelength range can help to determine accurate N(\oi) by detecting weaker \oi\ transitions \citep[e.g.,][]{pettini02,dessauges03,petitjean08}. In cases where oxygen abundance is difficult to measure, sulphur and silicon are used as representative of $\alpha$-capture elements \citep[see][]{centurion03}. After the O-burning process, the core of the massive stars is composed primarily of silicon and sulphur. Both elements track each other over a wide range of metallicities in Galactic stars \citep[e.g.,][]{chen02,nissen02} and in dwarf and BCD galaxies \citep[e.g.,][]{skillman94,izotov99}. Sulphur is also a non-refractory element and is therefore preferred over silicon. Si is mildly affected by depletion of the metal out of the gas phase \citep{vladilo11}, but has the advantage of being widely accessible in quasar absorbers. We note that both \sii\ and \siii\ are the dominant ions in \hi\ neutral gas, but can also be present in diffuse \hii\ ionised gas, in which case they could alter the interpretation of the [N/S] and [N/Si] measurements.

\begin{figure*}
\centering
{\includegraphics[width=\columnwidth,clip=]{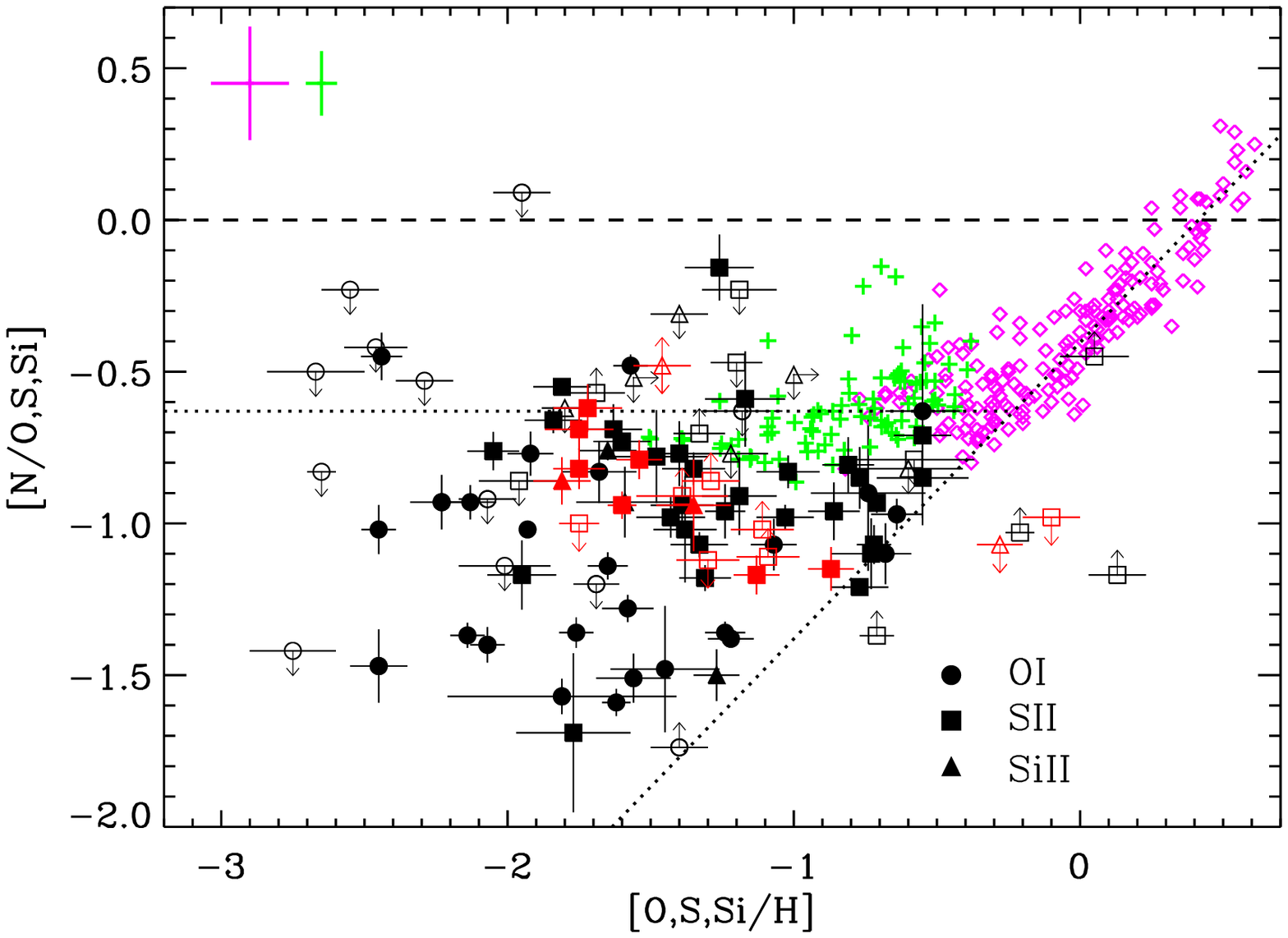}}
{\includegraphics[width=\columnwidth,clip=]{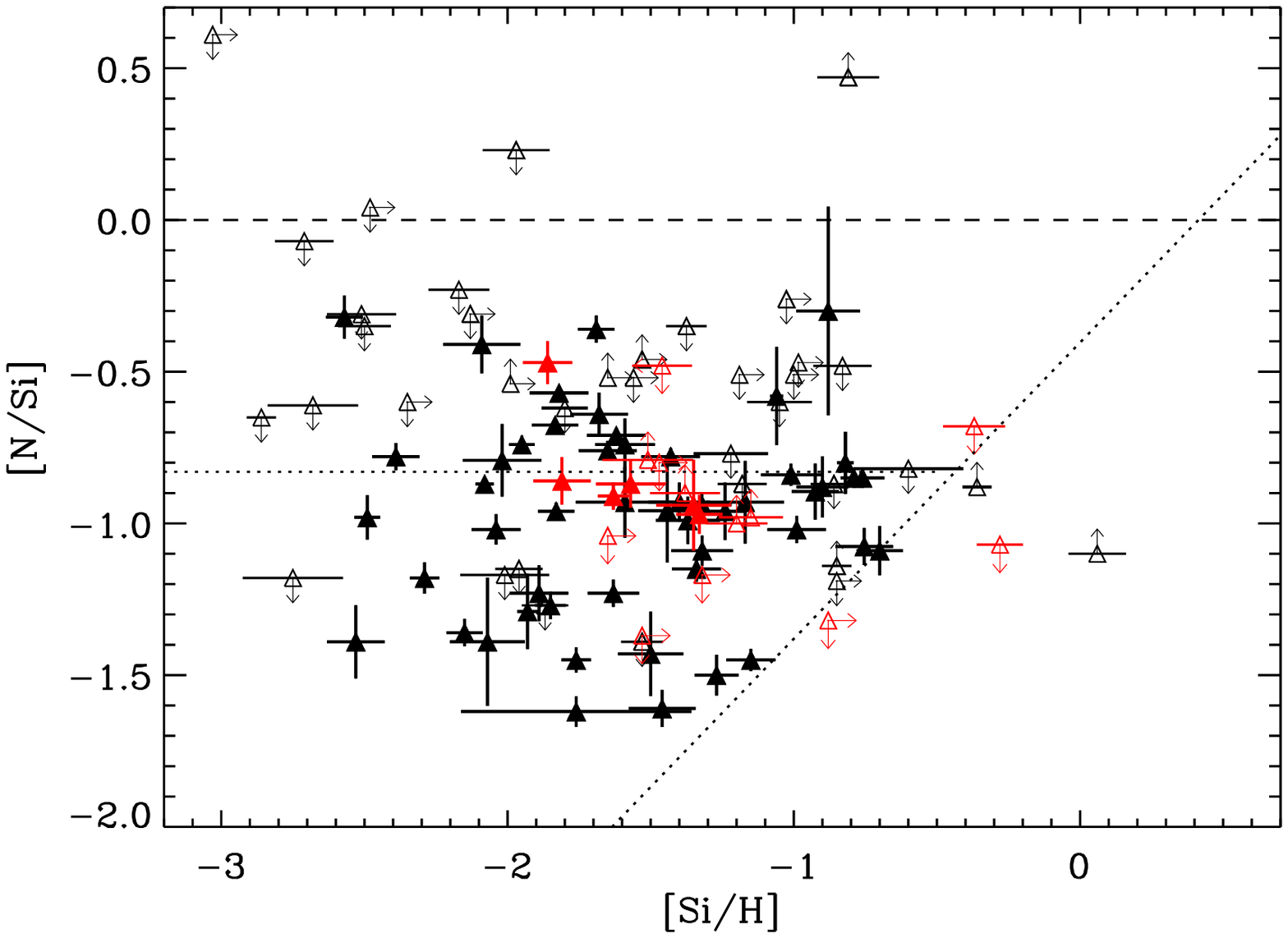}}
\caption{{\it Left panel:} The [N/$\alpha$] ratios against the $\alpha$-capture element metallicity. The circles, squares, and triangles indicate the oxygen, sulphur, and silicon abundances in DLAs/sub-DLAs, respectively. The filled symbols are for measurements while open ones are for upper or lower limits. The red data points are the new estimates presented in this work. The magenta diamonds and green pluses are measurements in the \hii\ regions of spiral \citep{vanzee98} and BCD galaxies \citep{izotov04} respectively. The magenta and green error bars on the top-left corner correspond to the average errorbars for the \hii\ regions of spiral and BCD galaxies, respectively. Dotted lines are empirical representations of the secondary and primary nitrogen production. The horizontal line (primary) is plotted at the mean value of [N/O] in BCD galaxies. The diagonal dotted line corresponds to the secondary N production observed in spiral \citep{vanzee98} and metal-poor emission-line galaxies \citep{nava06} and extrapolated to lower metallicities probed by quasar absorbers. The dashed line indicates the solar level. {\it Right panel:} Same as before but using only the $\alpha$-element silicon in DLAs/sub-DLAs. The primary N production (horizontal dotted line) is plotted at the mean [N/Si] value in DLAs/sub-DLAs.}
\label{NAOH}
\end{figure*}

 \subsection{The [N/$\alpha$] Bimodality}
In left panel of Fig. \ref{NAOH} and Fig. \ref{NHNSi}, the classical [N/$\alpha$] ratios versus $\alpha$-capture element metallicity as traced by [$\alpha$/H] and nitrogen abundance are shown for DLAs/sub-DLAs. There \oi, \sii\ and \siii\ are used as estimators of the $\alpha$-abundance. The right panels of both figures show the homogeneous set of measurements using \siii\ which is the $\alpha$-element most widely studied in DLAs. The use of Si instead of O and S does not significantly change the behaviour of [N/$\alpha$] ratios. This suggests that Si dust depletion is not large for the majority of the DLAs/sub-DLAs. We also compared our results with the measurements in the \hii\ regions of spiral \citep{vanzee98} and BCD galaxies \citep{izotov04}. Clearly, measurements in BCD galaxies indicating a constant ratio with metallicity, seem to favour a primary N production in these systems albeit with a large scatter. With the largest DLA/sub-DLA nitrogen abundance sample, it is evident that [N/$\alpha$] ratios are mainly spread over two regions. However, we gauge bimodality through various statistical tests.

\subsubsection{Testing Bimodality}
We quantify whether the [N/$\alpha$] sample is unimodal or bimodal. We first test whether [N/$\alpha$] sample is uniformly distributed or not. To test uniformity, we use $\chi^2$-test for uniform distribution and Ryan-Joiner normality (RJN) test to the unbinned entire sample (with and without limits). $\chi^2$-test suggested significant difference from the uniform distribution with 99\% probability. RJN-test rejects the hypothesis of normal distribution with $p$-value $<0.1$. This demonstrates departure from unimodal [N/$\alpha$] behaviour.

We then divided the [N/$\alpha$] sample into two groups: high-[N/$\alpha$]$>-1.2$ and low-[N/$\alpha$]$<-1.2$. This cutoff value is adopted throughout the analysis for defining both groups. We applied the Wilcoxon Rank Sum (RS) test, as well as the F-statistic test to the both groups to check bimodality. The RS-test resulted in a 100\% probability rejecting the hypothesis that the populations have the same distribution mean for both the samples with and without limits. The difference between both high and low [N/$\alpha$] populations is also evident in the cumulative distribution (see inset Fig. \ref{nahist}). The F-statistic test confirms that the two sample populations have significantly different variances. Considering the measurements only, the F-probability for the [N/$\alpha$] being bimodal is 88\%. The probability increases to 98\% if limits are included in the analysis. Hence these various statistical tests demonstrate with a high confidence level that the [N/$\alpha$] distribution is bimodal.



\begin{figure*}
\centering
{\includegraphics[width=\columnwidth,clip=]{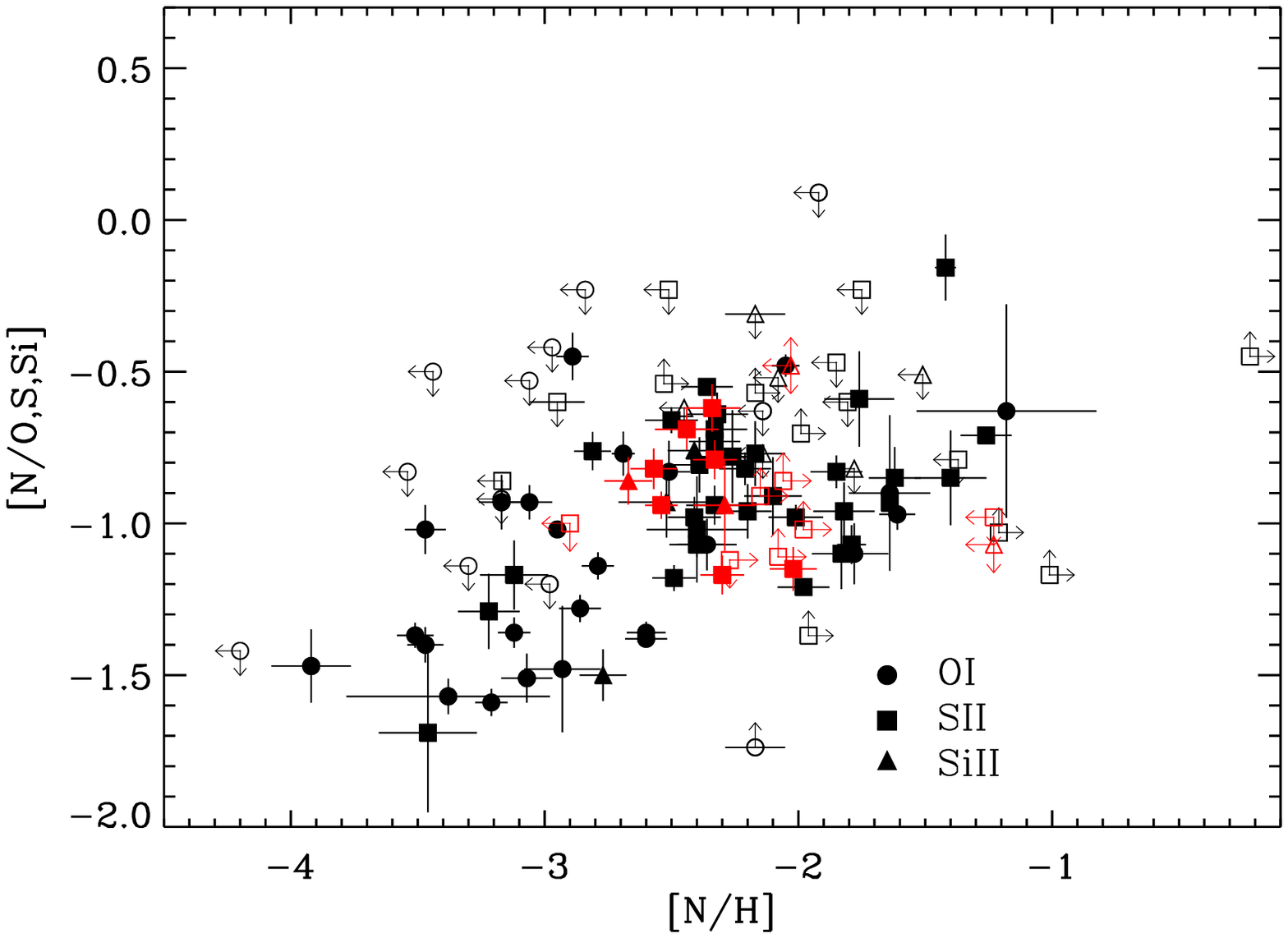}}
{\includegraphics[width=\columnwidth,clip=]{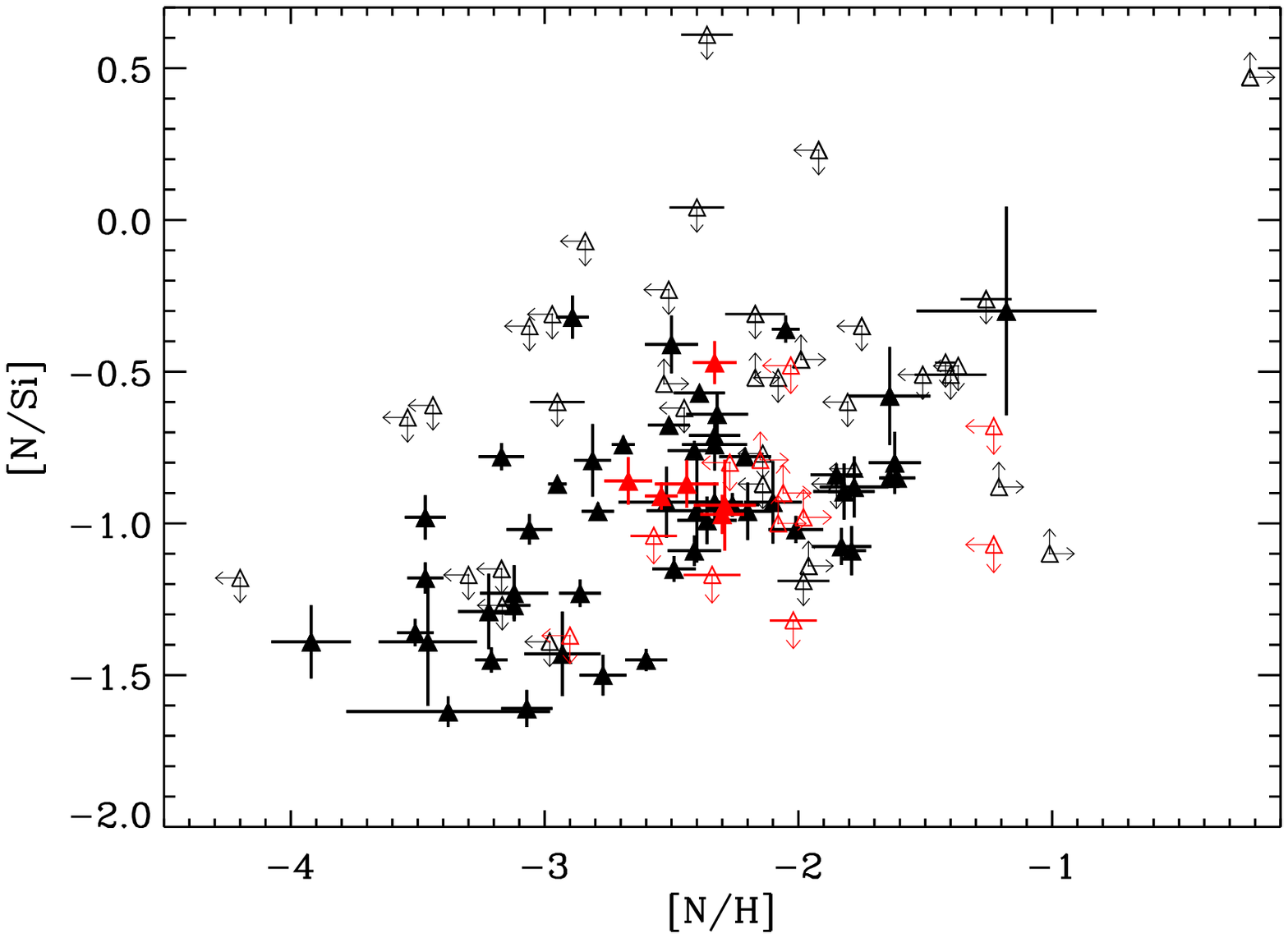}}
\caption{{\it Left panel:} The [N/$\alpha$] against the nitrogen abundance in DLAs/sub-DLAs. 
{\it Right panel:} same as before but using only the $\alpha$-element silicon.}
\label{NHNSi}
\end{figure*}

The low [N/$\alpha$] plateau appears at Nitrogen abundances below [N/H] $\simeq$$-2.5$ (see Fig. \ref{NHNSi}). Previously, there was a net separation between the two populations at [N/H] $\sim-2.8$--3.0 \citep[see][]{centurion03,molaro03}. With our three times larger sample, we find that there is a gradual transition between the groups. There are few systems (4 systems) below [N/H]$\simeq$$-2.5$ scattered in between the lower and upper plateaux and there are another 4 systems with very low-N abundance ([N/H]$\leq$$-3$) lying in the high [N/$\alpha$] plateau ([N/$\alpha > -1.2$).

The frequency distribution of the [N/$\alpha$] ratio is shown in Fig.~\ref{nahist}. The abundance ratios, [N/$\alpha$] and [N/Si] listed in Table~\ref{N:abund} are used to produce these histograms. We considered only the measurements for this analysis. Most of the systems appear in both histograms with the exception of 8 absorbers. Two peaks are clearly seen on both sides of [N/$\alpha$]$=-1.2$ although one of them has lower intensity. We fitted gaussian profiles to the binned distributions (bin size = 0.11) to materialise both groups. The adopted bin size is slightly larger than the mean error size (0.09) of the sample. We fitted single ($\chi^2/dof=142/13$) and double ($\chi^2/dof=31/10$) gaussian profiles to the entire sample. The best-fit result indicates that the data prefers a double gaussian fit. We find that the high [N/$\alpha$] cluster contains 77\% of DLAs/sub-DLAs (53 out of 69), showing a mean value of [N/$\alpha$] $= -0.85$ with a scatter of 0.20\,dex. The remaining 23\% DLAs/sub-DLAs (16 out of 69) belong to the low [N/$\alpha$] group, clustered around a mean value of [N/$\alpha$] $=-1.41$ with a slightly lower dispersion of 0.14\,dex. The results while including limits are: high-[N/$\alpha$] $=-0.84$ (standard deviation 0.27\,dex), and low-[N/$\alpha$] $=-1.37$ (standard deviation 0.14\,dex). Similar analysis for the [N/Si] ratios gives 77\% of the systems (44 out of 57) in the high [N/$\alpha$] cluster with [N/Si] $=-0.83$ with a standard deviation of 0.21\,dex and 23\% (13 out of 57) in the low [N/$\alpha$] cluster with [N/Si] $= -1.40$ with a scatter of 0.13\,dex. It should be noted that the majority of the [Si/H] and [O/H] and therefore [N/Si] and [N/O] values are consistent within 1$\sigma$. So the bimodal behaviour does not depend whether the metallicity is based on O or Si.

\citet{prochaska02} were first to report a possible bimodal distribution in the [N/$\alpha$] ratios of DLAs. Later \citet{centurion03} reported the existence of two groups: high [N/$\alpha$] sample (75\% of the systems) lying in a region comparable to the BCD galaxies at around [N/$\alpha$] $\simeq$$-0.87$ ($\pm$0.16\,dex), and low [N/$\alpha$] group (25\% of the systems) which is a factor of 5 lower in [N/$\alpha$] corresponding to [N/$\alpha$] $\sim$$-1.45$. These results were further confirmed by \citet{molaro03,molaro04,dodorico04,henry07} and \citet{pettini08}. \citet{petitjean08} found a less pronounced bimodality in their data using 13 measurements of [N/O]. The low [N/$\alpha$] measurements in DLAs have increased from 4 in 2003 to 16 with the extended sample presented here.

In Fig.~\ref{SiHNH} the [Si/H] ratio are plotted as a function of [N/H]. The two groups show a strong correlation with N tracing Si and with a clear offset between the two groups. The derived correlation coefficients are 0.94 and 0.90 for the low and high-N groups, respectively. The fact that N goes in lockstep with Si  is evidence of a primary-like behaviour of N in both groups. 

In spite of the enlargement of the present sample of nitrogen data, we do not find values of [N/$\alpha$] lower than $\sim$$-1.7$\, dex. The lack of such low values is not due a detection limitation, as one can see from the detection limits that we can attain with the present instrumentation. With VLT/UVES and Keck/HIRES, the detection limit for N is about [N/H] $\simeq$$-4.0$\,dex \citep[see][]{centurion03}; in fact, there is one measurement with [N/H] $\simeq$$-4.0$\,dex (DLA at \zabs=2.618 towards Q\,0913+0715) and a restrictive limit [N/H] $\leq$$-4.2$\,dex (DLA at \zabs =3.697 towards J\,0140-0839). Since there are many systems with [$\alpha$/H] $>$$-2.6$\,dex (see [$\alpha$/H] values in Table~\ref{N:abund}), we should be able to find systems with [N/$\alpha$] $\leq$$-1.7$\,dex. The lack of these systems suggests the existence of an intrinsic floor of  [N/$\alpha$] values. More observations of metal-poor systems \citep{pettini08,petitjean08,cooke11,dutta14} may provide more robust evidence on the existence of this floor of  [N/$\alpha$] values. Note that the DLA towards Q\,1425$+$6039 has near solar [N/$\alpha$] ratio.

\subsection{The high N/$\alpha$ plateau}
 
The high N/$\alpha$ ratios measured in the majority of DLAs/sub-DLAs are [N/$\alpha$] $=-0.85$ (standard deviation 0.20\,dex) slightly below the level of $-0.61$ (standard deviation 0.14\,dex) measured in BCGs \citep{izotov99,izotov04} and \hii\ regions of dwarf-irregular galaxies \citep{vanzee06}. These DLA values identify a plateau  which extends to much lower metallicities thus strengthening  the need for a primary source for nitrogen occurring at low metallicities. The [N/O] plateau in the low metallicity tail of the BCGs is interpreted as due to the primary production of N by long-lived IMS (4 $\leq$ $M/M_\odot$ $\leq$ 8) during the AGB phase, where dredged-up episodes bring freshly synthesized C into H-burning upper shells where is transformed into N by the CNO cycle \citep{henry00,marigo01,chiappini03}. The similar [N/$\alpha$] value of DLA and BCG suggests a similar chemical enrichment in both populations.

The delayed release of N into the ISM with respect to O which is ejected by short-lived massive stars when they exploded as Type\,II supernovae might explain the scatter observed in the BCG. \citet{pettini02} and more recently  \citet{cooke11,dutta14}, propose that the same model of a delayed release of N could also be responsible for the whole DLA data points when a large scatter is considered instead of a bimodal distribution. However, the clustering around two values would remain unexplained and the delayed model is unlikely capable of producing a factor 3 increase in the dispersion of the [N/O] values. The DLA galaxies show low star-formation rates and the time required to make a substantial amount of O  becomes comparable to the lag time of the IMS ($\sim$250\,Myrs) to deliver primary N in the ISM \citep{henry07}. This should lead to an increase of O and N almost in lockstep producing a scatter that should be comparable or even smaller than that observed in the BCGs. Moreover, to produce a very low [N/O] ratio we require a significant O enrichment combined with little or no N production. In DLAs, the low [N/O] values are observed preferentially for relatively low values of both N and O as it is shown in  Fig.~\ref{SiHNH} where Si is used as a proxy for O.

\subsection{The low N/$\alpha$ plateau}
\begin{figure}
\begin{center}
{\includegraphics[width=\columnwidth,clip=]{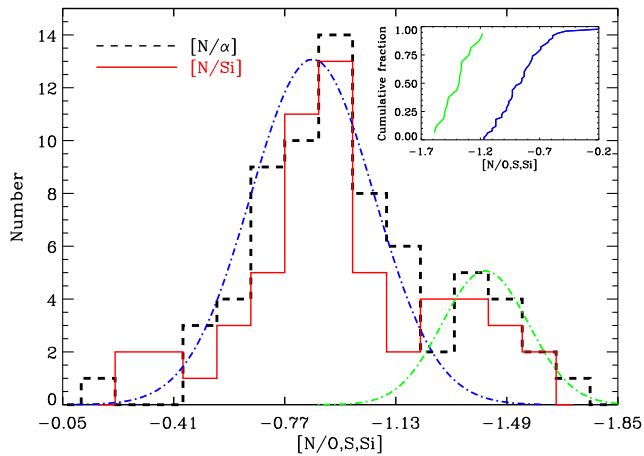}}
\caption{Histograms of [N/$\alpha$] abundance ratios for the 68 DLAs and sub-DLAs in the sample. The adopted bin size is 0.11. The abundance ratios [N/$\alpha$] and [N/Si] listed in Table~\ref{N:abund} are used to produce these histograms. Therefore, most of the systems appear in both histograms with the exception of 8 absorbers. To illustrate bimodality, gaussian fits for high-[N/$\alpha$] (blue dashed-dotted line) and low-[N/$\alpha$] (green dashed-dotted line) distributions are plotted. {\it Inset}: The cumulative distributions for both distributions are shown.}
\label{nahist}
\end{center}
\end{figure}

\citet{centurion03} and \citet{molaro03} suggested that  this plateau could be the result of  primary production of N by young massive stars. If massive stars were able to produce primary N then we might expect to see a floor on the N/O ratios since both elements are ejected into the ISM at the same time, and the N/O values can only increase at a later time with the contribution of the IMS. In this framework, low [N/$\alpha$] DLAs would be young objects caught before the ejection of primary N by the IMS. The primary behaviour of the low [N/$\alpha$] ratios shown in Figs. \ref{NAOH}, \ref{NHNSi} and \ref{nahist} and the lower scatter of these ratios are in line with this scenario. If massive stars produce primary N in the systems which lie in the lower plateau, then it is reasonable to expect that these systems should evolve, moving towards the upper plateau by increasing the N-abundance and then the N/O ratios, due to the delayed contribution of the IMS \citep{henry00}. In this simplified scenario it is possible that DLAs in the transition zone are caught during the secondary N enrichment by IMS.

\begin{figure}
\begin{center}
{\includegraphics[width=\columnwidth,clip=]{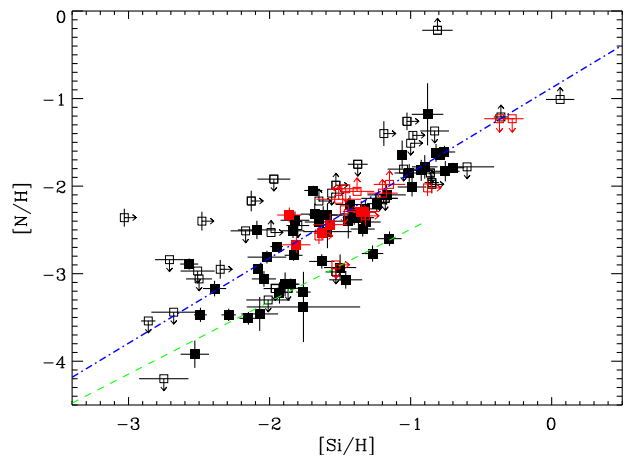}}
\caption{The Si abundance versus N abundance in DLAs/sub-DLAs.
The data appear to be divided in two groups for which the best-fits are indicated by green ($\beta=0.83\pm0.09$) and blue ($\beta=0.94\pm0.03$) dashed lines, respectively.}
\label{SiHNH}
\end{center}
\end{figure}

The models of nucleosynthesis of zero-metalllicity stars (Pop\,III) of \citet{heger10} and \citet{limongi12} can produce primary N at the base of the hydrogen envelope. When carbon and hydrogen get mixed as a consequence of convective overshoot, carbon is easily converted into nitrogen. This occurs for stars with masses greater than 20--25$M_\odot$, it goes trough a minimum at about 40$M_\odot$ to rise again for higher mass stars up to 80$M_\odot$ in \citet{heger10} but not in \citet{limongi12}. The yields in these models are quite high producing peak ratio of [N/O]$\sim -0.3$, i .e. in stars with 25$M_\odot$ but can reproduce the observed range when integrated over a Salpeter IMF. However, the yields are quite uncertain by one order of magnitude and are rather sensitive to rotation. 
\citet{cooke11} find (see their Fig.\,14) that either models of zero-metallicity Pop\,III stars with 10$<$M/M$_\odot$$< $100, by \citet{heger10} or Pop\,II stars with metallicity $Z$$=$$0.05 Z_\odot$ and masses in the range 13--35 M$_\odot$ by \citet{chieffi04} can reproduce the low [N/O] ratios observed. Both models are able to reproduce the observed $<$[Si/O]$>$$\approx$$-0.04$ and $<$[Fe/O]$>$$\approx$$-0.38$ for the systems in the lower plateau. However, in this latter case the N production should be secondary which is not in line with the fact that N grows in lockstep with O. In \citet{heger10} model, a problem is to understand why N follows Si. In fact the initial [N/O] value left by Pop\,III stars should be followed by an initial decrease in the [N/O] ratio due to O, but not N, production by the second generation of stars followed by the N secondary production by next stellar generations. This complex pattern is not apparent in Fig.~\ref{SiHNH} where N seems to increase in lockstep with O as a pure primary element.
A possibility of explaining primary N production due to metal-poor IMS and massive stars is by introducing stellar rotation \citep{meynet02,chiappini06}. Primary N is produced during the He-burning phase by rotational diffusion of C into the H-burning shell. The output of the models show dependence on the stellar lifetimes, star-formation efficiencies, and surface velocities.

The different state of evolution of the two groups should produce different [$\alpha$/Fe-peak-elements] enhancements: qualitatively one would expect a lack of [$\alpha$/Fe] enhancement for the DLAs with high [N/$\alpha$] ratios, and [$\alpha$/Fe] enhanced ratios for the DLAs with low [N/$\alpha$] ratios.
In fact if primary N in the lower plateau is produced by massive stars at the same time as O, the delayed ejection of Fe mainly produced by Type\,Ia supernovae explosions of low and IMS will cause enhanced O/Fe ratios in the lower plateau compared to those in the upper one. Both the high and low [N/$\alpha$] plateaux show enhanced [Si/Fe] ratios with values between 0.3 and 0.4, which are rather common in DLAs. However, the systems in the upper plateau have solar [Si/Zn] ratios indicating that Fe is depleted onto dust and the observed [$\alpha$/Fe-peak-elements] enhancement is not real. The same could be true also for the enhanced [$\alpha$/Fe] ratios observed for the systems in the lower plateau. However, our expectation is that they should be less evolved systems having a lower amount of dust if any. Unfortunately, the lack of Zn information for the low [N/$\alpha$] group prevents us to make a conclusive test.

The sub-DLAs that are present in our compilation (20 out of 108 systems) belong, for the major part, to the low [N/$\alpha$] group. The potential presence of ionisation effects that may alter the measured abundances is a reason of concern for these systems of relatively low N(\hi). Should ionisation effects be important, we would expect a correlation between N(\hi) and the [N/$\alpha$] ratios, with $\alpha$ = \siii\ or \sii\, the ions which could be affected by the presence of \hii\ gas. Instead, we do not find this type of correlation, with a correlation coefficient of $-0.03$, in agreement with findings of \citet{petitjean08}. A further indication for lack of ionisation effects is that the \siii/\oi\ or \sii/\oi\ ratios, when available in sub-DLAs, are not enhanced (see Table \ref{N:abund}) as it would be expected from the presence of \hii\ gas. The lack of ionisation effects may be due to the fact that 15 out 20 of the sub-DLAs of our sample have relatively high column densities, log N(\hi) $>$$20$\,dex. All together, we believe that ionisation effects do not affect the estimate of the N/S and N/Si ratios in our sample of sub-DLAs. Therefore, the position of the sub-DLAs on the lower plateau can be explained by the fact that sub-DLAs are on average younger and less-massive \citep{kulkarni07}, still in the earliest stages of chemical evolution and/or experience a slower evolution.

\begin{figure}
\begin{center}
{\includegraphics[width=\columnwidth,clip=]{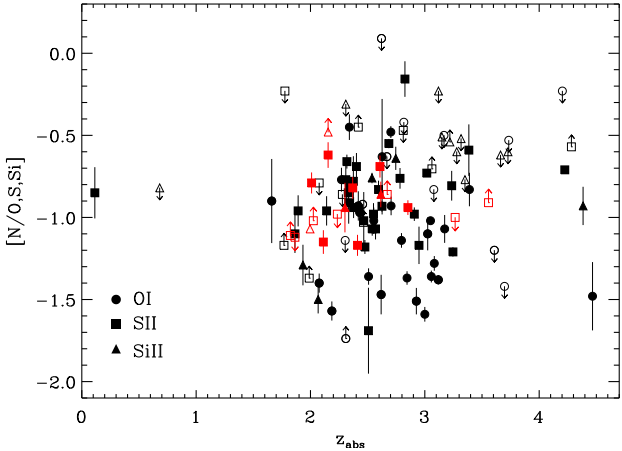}}
\caption{The [N/$\alpha$] ratios in DLAs/sub-DLAs as a function of redshift, \zabs .}
\label{zabsNSi}
\end{center}
\end{figure}

\subsection{The [N/$\alpha$] redshift evolution}
The [N/$\alpha$] results have important implications concerning the epoch of formation of the DLA galaxies of our sample. The evolution with cosmic time of the nitrogen abundances can be appreciated in Fig.~\ref{zabsNSi}, where we plot the [N/$\alpha$] values as a function of absorption redshift. In this figure one can see that the [N/$\alpha$] values show a large spread. Systems with high and low [N/$\alpha$] ratios do not occupy a specific region of the diagram, but are instead observed at all values of redshift covered by our sample. As discussed in the previous sections, low [N/$\alpha$] values are expected in relatively young objects where the IMS did not yet have time to evolve and contribute to the nitrogen enrichment, while high [N/$\alpha$] values are observed where these stars evolved and contributed significantly to the bulk of N. As a consequence, all objects with low [N/Si] abundance are young independently from their redshift. Here, old and young refers to the characteristic timescale of the main N production by the IMS, which can be of the order of 250--500 Myrs \citep{henry00}, or even longer if rotation plays an important role \citep{meynet02}. The fact that these DLA absorbers are observed  at all epochs  indicate a continuous formation of DLAs rather than a specific epoch for their formation.

In our sample there are 3 objects with \zabs $>$4 that yield stringent limits on the epoch of formation of their host galaxies: the DLAs towards J\,0307-4945 at \zabs=4.466, the one at redshift \zabs=4.383 observed toward B\,1202-074 and the one at \zabs=4.383 towards J\,1443+2724. High [N/$\alpha$] values are observed in the cases of B\,1202-074 (\zabs=4.384) and J\,1443+2724 (\zabs=4.224) which  correspond to relatively old systems. Adopting a scale time of 500\,Myrs for full N production by the IMS, the onset of star formation in these two DLAs took place already at redshift $> 6$  pushing back considerably the epoch of star-formation \citep[see][]{molaro04,dodorico04}.


\section{Conclusions}

Studies of nitrogen in DLAs/sub-DLAs provide important clues of the earlier stages of galactic chemical evolution. The N abundance determination in sites of low and high metallicities plays an important role in understanding the nucleosynthetic origin of nitrogen. Previous studies have shown that DLAs follow a primary N production at low-metallicities, consistent with the BCDs and dwarf irregular galaxies measurements. In addition, another small population exists, lying well below the primary plateau. We here attempted to test the presence of this population with an extended sample. The results from this work are summarised as follows:

\begin{itemize}

\item We present 9 new \ni\ measurements, 5 upper and 4 lower limits in DLAs/sub-DLAs mostly from the EUADP sample \citep{zafar13,zafar13b}. We combine these data with a careful reappraisal of literature high-resolution measurements published to date, making a total sample of 70 \ni\ measurements and 38 upper and lower limits. This is the largest nitrogen abundance sample studied so far.

\item We derived [N/$\alpha$] abundance ratios of our sample. We performed $\chi^2$ and Ryan-Joiner normality tests, indicating that the [N/$\alpha$] distribution is not unimodal. Furthermore, considering two populations with [N/$\alpha$] larger or smaller than $-1.2$ the RS and F-statistic tests demonstrate that they are not drawn from the same distribution and have significant differences. Three-quarter of the DLAs are in the upper group and show a mean value of [N/$\alpha$] $=$ $-0.85$ with a scatter of 0.20\,dex and one-quarter in the lower group shows ratios clustered at [N/$\alpha$] $=$ $=1.41$ with a slightly lower dispersion of 0.14\,dex.
\item The low [N/$\alpha$] plateau appears at Nitrogen abundances below [N/H] $\simeq-2.5$. The transition between the groups is gradual and there are 8 systems scattered in between the groups. This is consistent with our hypothesis because we expect that the transition region is populated by evolving systems where nitrogen produced by the IMS is enriching the ISM. This process happens very rapidly so the observed transiting systems are rarer.
 
\item The high [N/$\alpha$] plateau is consistent with the results found in the BCD \citep{izotov04} and dwarf irregular galaxies \citep{vanzee06}. This plateau supports the idea that IMS (4 $\leq$ $M/M_\odot$ $\leq$ 8) during their AGB phases are the dominant sites of primary N production. In such case the N goes in lockstep with O, resulting in a constant [N/O] ratio. The larger dispersion is consistent with the delayed delivery of primary N in the ISM with respect to O.

\item The [N/$\alpha$] ratios in the low plateau are the lowest ever measured in any astrophysical site. Furthermore, we confirm the existence of a [N/$\alpha$] floor at $\sim$$-1.4$ which is not due to observational bias. We interpret these low [N/$\alpha$] ratios as due to primary N production by fast rotating young massive stars. According to nucleosynthesis models of massive stars, the low [N/$\alpha$] plateau could be the signature of Pop\,III stars, in the earliest episodes of N enrichment. It is reasonable to expect that the low [N/$\alpha$] systems should evolve moving towards the upper plateau by increasing their N abundance and hence the [N/$\alpha$] ratios.

\item There are only two systems around [N/H]$\sim$$-4$. This is due to the lack of metal-poor DLAs with low-N abundances. To further probe the low [N/$\alpha$] population, it would be important to extend the observations of VMP systems. This will provide potential clues about the nitrogen enrichment histories in these sites.

\end{itemize}

\section*{Acknowledgements}
We would like to thank the ESO staff for making the UVES Advanced Data Products available to the community. We are thankful to the anonymous referee for his/her constructive report. CP would like to thanks the BINGO! (`history of Baryons: INtergalactic medium/Galaxies cO-evolution') project by the Agence Nationale de la Recherche (ANR) under the allocation ANR-08-BLAN-0316-01. We are thankful to Ryan Cooke for helpful comments and for providing revised abundance measurements of his data. M.C., V.D. and G.V. acknowledge support from the PRIN INAF ``The 1 Billion Year Old Universe: Probing Primordial Galaxies and the Intergalactic Medium at the Edge of Reionization''. P. M. acknowledges the discussions with Marco Limongi and Alessandro Chieffi and the international team \#272 lead by C. M. Coppola ``EUROPA - Early Universe: Research on Plasma Astrochemistry'' at ISSI (International Space Science Institute) in Bern.

\begin{table*}
\begin{minipage}{180mm}
\begin{center}
\setlength{\tabcolsep}{2.5pt}
\caption{N and $\alpha$-element abundance measurements in DLAs and sub-DLAs. The new estimates presented in this work are shown in bold-face. The column $\alpha$ refers to the choice of $\alpha$ element for the determination of the [$\alpha$/H] and [N/$\alpha$] ratio.}
\label{N:abund}
\begin{tabular}{l c c c c c c c c c c}
\hline\hline                       
QSO & $z_{abs}$ &  log $N$(\hi) & log $N$(\ni) & [N/H]  & [Si/H]	& [N/Si] & $\alpha$  & [$\alpha$/H] & [N/$\alpha$] & Ref. \\
	 & & cm$^{-2}$ 	& cm$^{-2}$ &  \\ 
\hline
Q\,0000-2620 	& 3.390 	& $21.41\pm0.08$ 	& $14.73\pm0.02$ 	& $-2.51\pm0.08$ 		& $-1.83\pm0.08$ 		& $-0.68\pm0.02$ 	& O  & $-1.68\pm0.13$   & $-0.83\pm0.10$ 	& 1 \\
QXO\,0001 		& 3.000		& $20.70\pm0.05$ 	& $13.32\pm0.04$ 	& $-3.21\pm0.06$ 		& $-1.76\pm0.05$ 		& $-1.45\pm0.04$ 	& O  & $-1.62\pm0.05$   & $-1.59\pm0.04$ 	& 2 \\	
J\,0003-2323$^\dagger$ 	& 2.187 	& $19.60\pm0.40$ 	& $12.05\pm0.03$ 	& $-3.38\pm0.40$ 		& $-1.76\pm0.40$ 		& $-1.62\pm0.05$ 	& O  & $-1.81\pm0.40$   & $-1.57\pm0.06$ 	& 3 \\
{\bf LBQS\,0010-0012} & 2.025 	& $20.95\pm0.10$  	& $>14.80$ 		& $>-1.98$ 			& $-1.15\pm0.11$ 		& $>-0.98$ 	 	& S  & $-1.11\pm0.11$   & $>-1.02$ 	 	& 4, 5 \\	
B\,0027-1836 	& 2.402 	& $21.75\pm0.10$ 	& $15.25\pm0.08$ 	& $-2.33\pm0.13$ 		& $-1.59\pm0.10$ 		& $-0.74\pm0.09$ 	& S  & $-1.64\pm0.10$   & $-0.69\pm0.08$ 	& 6 \\
J\,0035-0918 	& 2.340 	& $20.43\pm0.04$ 	& $13.37\pm0.05$ 	& $-2.89\pm0.06$ 		& $-2.57\pm0.06$ 		& $-0.32\pm0.07$ 	& O  & $-2.44\pm0.07$   & $-0.45\pm0.08$ 	& 7 \\
{\bf B\,0058-292}	& 2.671		&$21.10\pm0.10$ 	& $>14.87$ 		& $>-2.06$ 			& $-1.38\pm0.12$ 		& $>-0.89$ 	 	& S  & $-1.29\pm0.10$   & $>-0.86$ 	 	& 4, 5  \\
B\,0100+1300 	& 2.309 	& $21.37\pm0.08$ 	& $15.03\pm0.09$ 	&  $-2.17\pm0.12$  		& $>-2.11$ 			& $<-0.31$ 	 	& S  & $-1.40\pm0.10$   & $-0.77\pm0.11$ 	& 8, 9 \\
{\bf J\,0105-1846} & 2.369 	& $21.00\pm0.08$ 	& $14.26\pm0.04$ 	& $-2.57\pm0.09$ 		& $>-1.65$ 		& $<-1.04$ 	& S  & $-1.75\pm0.09$  	& $-0.82\pm0.07$ 	& 4  \\
J\,0105-1846 	& 2.926 	& $20.00\pm0.10 $ 	& $12.76\pm0.01$ 	& $-3.07\pm0.10$  		& $-1.46\pm0.12$ 		& $-1.61\pm0.06$ 	& O  & $-1.56\pm0.13$   & $-1.51\pm0.08$ 	& 10 \\
Q\,0112-306	   &  2.418 	& $20.50\pm0.08$ 	& $13.16\pm0.14$ 	& $-3.17\pm0.09$  		& $-2.39\pm0.08$ 		& $-0.78\pm0.04$ 	& O  & $-2.24\pm0.11$   & $-0.93\pm0.09$ 	& 11 \\
{\bf B\,0122-005} 	& 2.010 	& $20.04\pm0.07$ 	& $13.54\pm0.05$ 	& $-2.33\pm0.09$ 		& $-1.86\pm0.09$ 		& $-0.47\pm0.07$ 	& S  & $-1.54\pm0.08$   & $-0.79\pm0.06$ 	& 4 \\
J\,0127-0045	& 3.727	& $21.15\pm0.10$ &	$14.03\pm0.04$ & $-2.95\pm0.11$ & 	$>-2.35$ & 	$<-0.60$ & Si & $>-2.35$ & 	$<-0.60$ & 12 \\
J\,0140-0839 	& 3.697 	& $20.75\pm0.15$ 	& $\leq12.38$ 		& $\leq-4.20$ 			& $-2.75\pm0.17$ 		& $\leq-1.18$ 	 	& O  & $-2.75\pm0.15$   & $\leq-1.42$ 	 	& 13 \\
Q\,0151+048 	& 1.934 	& $20.34\pm0.02$ 	& $12.95\pm0.12$ 	& $-3.22\pm0.12$ 		& $-1.93\pm0.04$ 		& $-1.29\pm0.12$ 	& Si  & $-1.93\pm0.04$ 	& $-1.29\pm0.12$ 	    & 14 \\
B\,0201+113 	& 3.385 	& $21.26\pm0.08$ 	& $15.33\pm0.11$ 	& $-1.76\pm0.14$  		& $\cdots$ 			& $\cdots$ 	 	& S  & $-1.17\pm0.14$   & $-0.59\pm0.16$ 	& 15 \\
Q\,0201+365 	& 2.463 	& $20.38\pm0.05$ 	& $>15.00$ 		& $>-1.21$ 			& $-0.36\pm0.05$ 		& $>-0.88$ 	 	& S  & $-0.21\pm0.05$   & $>-1.03$ 	 	& 2 \\	
J\,0234-0751 & 2.318  	& $20.90\pm0.10$ 	& $14.23\pm0.03$ 	& $-2.50\pm0.10$ 		& $-2.09\pm0.13$ 		& $-0.41\pm0.09$ 	& S  & $-1.84\pm0.10$   & $-0.66\pm 0.04$ 	&  16  \\
J\,0307-4945 	& 4.466 	& $20.67\pm0.09$ 	& $13.57\pm0.12$ 	& $-2.93\pm0.15$ 		& $-1.50\pm0.11$ 		& $-1.43\pm0.14$ 	& O  & $-1.45\pm0.19$   & $-1.48\pm0.21$ 	& 17 \\
J\,0311-1722 	& 3.734		& $20.30\pm0.06$ 	&  $\leq13.07$ 		& $\leq-3.06$ 			& $-2.50\pm0.09$ 		& $\leq-0.35$ 	 	& O  & $-2.29\pm0.10$   & $\leq-0.53$ 	 	& 6 \\
{\bf J\,0332-4455} & 2.411 	& $20.15\pm0.07$  	& $13.68\pm0.05$ 	& $-2.30\pm0.09$ 		& $-1.33\pm0.08$ 		& $-0.97\pm0.06$ 	& S  & $-1.13\pm0.08$   & $-1.17\pm0.06$ 	& 4 \\
{\bf Q\,0334-1612} & 3.557 & $21.12\pm0.15$ & $>14.80$ & $>-2.15$ & $-1.51\pm0.16$ & $>-0.79$ & S & $-1.39\pm0.16$ & $>-0.91$ & 4 \\
B\,0336-017 	& 3.062 	& $21.20\pm0.09$ 	& $>15.04$ 		& $>-1.99$ 			& $>-1.53$ 			& $>-0.70$ 	 	& S  & $-1.33\pm0.09$   & $>-0.70$ 	 	& 2 \\
B\,0347-383 	& 3.025 	& $20.63\pm0.09$ 	& $14.68\pm0.10$ 	& $-1.78\pm0.13$  		& $-0.90\pm0.09$ 		& $-0.88\pm0.10$ 	& O  & $-0.68\pm0.09$   & $-1.10\pm0.10$ 	& 18, 19 \\
J\,0407-4410 	& 2.551 	& $21.13\pm0.10$ 	&  $14.55\pm0.03$ 	& $-2.41\pm0.10$  		& $-1.32\pm0.11$ 		& $-1.09\pm0.05$ 	& S  & $-1.43\pm0.12$   & $-0.98\pm0.07$ 	& 19, 20 \\
J\,0407-4410 	& 2.595 	& $21.09\pm0.10$ 	& $15.07\pm0.02$ 	& $-1.85\pm0.10$ 		& $-1.01\pm0.10$ 		& $-0.84\pm0.04$ 	& S  & $-1.02\pm0.11$   & $-0.83\pm0.05$ 	& 19, 20 \\
J\,0407-4410 	& 2.621 	& $20.45\pm0.10$ 	& $<14.36$ 		& $<-1.92$ 			& $-1.97\pm0.12$ 		& $<0.23$ 	 	& O  & $-1.95\pm0.10$   & $<0.09$ 	 	& 19, 20 \\
{\bf B\,0432-440}	& 2.302		& $20.95\pm0.10$ 	& $14.49\pm0.07$ 	& $-2.29\pm0.12$ 		& $-1.35\pm0.13$ 	 	& $-0.94\pm0.15$ 	& Si & $-1.35\pm0.13$   & $-0.94\pm0.15$   & 4 \\
B\,0450-1310B 	& 2.067 	& $20.50\pm0.07$ 	&  $13.56\pm0.06$ 	& $-2.77\pm0.09$ 		& $-1.27\pm0.08$ 	 	& $-1.50\pm0.07$ 	& Si & $-1.27\pm0.08$   & $-1.50\pm0.07$ 	& 21 \\
B\,0528-2505 	& 2.141 	& $20.95\pm0.05$ 	&  $14.58\pm0.08$ 	& $-2.20\pm0.09$ 		& $-1.24\pm0.07$ 		& $-0.96\pm0.09$ 	& S  & $-1.24\pm0.07$   & $-0.96\pm0.09$ 	& 22, 23 \\
B\,0528-2505 	& 2.811 	& $21.35\pm0.07$ 	& $<15.33$ 		& $<-1.85$ 			& $-0.86\pm0.08$ 		& $<-0.87$ 	 	& S  & $-1.20\pm0.09$   & $<-0.47$ 	 	& 23, 24 \\
HS\,0741+4741 	& 3.017 	& $20.48\pm0.10$ 	& $13.98\pm0.01$ 	& $-2.33\pm0.10$ 		& $-1.62\pm0.10$ 		& $-0.71\pm0.02$ 	& O  & $-1.60\pm0.10$   & $-0.73\pm0.02$ 	& 2 \\
FJ\,0812+32 	& 2.626 	& $21.35\pm0.10$ 	&  $16.00\pm0.34$ 		& $-1.18\pm0.35$ 	& $-0.88\pm0.11$ 		& $-0.30\pm0.34$ 	 	& O  & $-0.55\pm0.13$   & $-0.63\pm0.35$ 	 	& 25 \\
J\,0831+3358 	& 2.304 	& $20.25\pm0.15$ 	&  $\leq12.78$ 		& $\leq-3.30$ 			& $-2.01\pm0.16$ 		& $\leq-1.17$ 	 	& O  & $-2.01\pm0.16$   & $\leq-1.14$ 	 	& 26 \\
{\bf B\,0841+129}	& 1.864 	& $21.00\pm0.10$  	& $<14.56$ 		& $<-2.27$ 			& $>-1.47$ 			& $<-0.80$ 	 	& S  & $-1.30\pm0.11$   & $<-1.12$ 	 	& 4, 27 \\	
B\,0841+129	& 2.375 	& $21.05\pm0.10$ 	& $14.62\pm0.03$ 	& $-2.26\pm0.10$ 		& $-1.32\pm0.10$ 		& $-0.94\pm0.04$ 	& S  & $-1.48\pm0.18$   & $-0.78\pm0.15$ 	& 8, 22 \\
B\,0841+129	& 2.476 	& $20.78\pm0.08$ 	& $14.12\pm0.03$ 	& $-2.49\pm0.09$ 		& $-1.34\pm0.09$ 		& $-1.15\pm0.04$ 	& S  & $-1.31\pm0.09$   & $-1.18\pm0.04$ 	& 8, 9, 22 \\
J\,0900+4215 	& 3.246 	& $20.30\pm0.10$ 	& $14.15\pm0.02$ 	& $-1.98\pm 0.10$ 		& $>-0.85$ 			& $<-1.19$ 	 	& S  & $-0.77\pm0.10$   & $-1.21\pm0.03$ 	& 28 \\
Q\,0913+0715 	& 2.618	& $20.35\pm0.10$ 	& $12.26\pm0.12$ 	& $-3.92\pm0.16$ 		& $-2.53\pm0.10$ 		& $-1.39\pm0.12$ 	& O  & $-2.45\pm0.10$   & $-1.47\pm0.12$ 	& 11, 29, 30 \\
Q\,0930+2858 	& 3.235	& $20.30\pm0.10$ 	& $13.74\pm0.01$ 	& $-2.39\pm0.10$ 		& $-1.82\pm0.10$ 		& $-0.57\pm0.02$ 	& S  & $-1.58\pm0.13$   & $-0.81\pm0.09$ 	& 2 \\
J\,0953-0504 	& 4.203 	& $20.55\pm0.10$ 	& $<13.54$ 		& $<-2.84$ 			& $-2.71\pm0.10$ 		& $<-0.07$ 	 	& O  & $-2.55\pm0.10$    & $<-0.23$ 	 	& 16 \\
J\,0953+5230 	& 1.768 	& $20.10\pm0.10$ 	& $>14.92$ 		& $>-1.01$ 			& $0.06\pm0.10$ 		& $>-1.10$ 	 	& S  & $0.13\pm0.10$    & $>-1.17$ 	 	& 31 \\
PSS\,0957+33 	& 3.279 	& $20.42\pm0.10$ 	& $<14.44$ 		& $<-1.81$ 			& $-1.05\pm0.11$ 		& $<-0.60$ 	 	& Si  & $-1.05\pm0.11$ 		& $<-0.60$ 	    	& 12, 18 \\
J\,1001+0343 	& 3.078		& $20.21\pm0.05$ 	& $\leq12.50$ 		& $\leq-3.54$ 		& $-2.86\pm0.05$ 		& $\leq-0.65$ 	 	& O  & $-2.65\pm0.05$   & $\leq-0.83$ 	 	& 26 \\
J\,1004+0018 	& 2.540 	& $21.30\pm 0.10$ 	& $14.73\pm0.04$ 	& $-2.40\pm0.11$ 		& $>-2.48$ 			& $<-0.04$ 	 	& S  & $-1.33\pm0.10$   & $-1.07\pm0.04$ 	&  16 \\
J\,1004+0018 	& 2.685 	& $21.39\pm 0.10$ 	& $14.86\pm0.02$ 	& $-2.36\pm0.10$ 		& $>-3.03$ 			& $<0.61$ 	 	& S  & $-1.81\pm0.10$   & $-0.55\pm0.03$ 	& 16  \\
J\,1004+0018 	& 2.746 	& $19.84\pm 0.10$ 	& $13.35\pm0.07$ 	& $-2.32\pm0.12$ 		& $-1.68\pm0.10$ 			& $-0.64\pm0.07$ 	 	& Si  & $-1.68\pm0.10$ 	& $-0.64\pm0.07$ 	&  16 \\
J\,1009+0713 	& 0.114 	& $20.68\pm 0.10$ 	& $15.11\pm0.10$ 	& $-1.40\pm0.14$ 		& $>-1.19$ 			& $<-0.51$ 	 	& S  & $-0.55\pm0.16$   & $-0.85\pm0.16$ 	& 32 \\
J\,1016+4040 	& 2.816 	& $19.90\pm0.11$ 	& $\leq12.76$ 		& $\leq-2.97$ 			& $-2.51\pm0.12$ 		& $\leq-0.31$ 	 	& O  & $-2.46\pm0.11$   & $\leq-0.42$ 	 	& 26 \\
BQ\,1021+3001 	& 2.949 	& $20.70\pm0.10$ 	& $13.41\pm0.09$ 	& $-3.12\pm0.13$ 		& $-1.89\pm0.10$ 		& $-1.23\pm0.09$ 	& S  & $-1.95\pm0.12$   & $-1.17\pm0.11$ 	& 25 \\
{\bf B\,1036-268 }	& 2.235 	& $19.96\pm0.09$ 	& $<14.57$ 		& $<-1.23$ 			& $-0.37\pm0.11$ 		& $<-0.68$ 	 	& S  & $-0.10\pm0.10$   & $<-0.98$ 	 	& 4 \\
J\,1037+0139 	& 2.705 	& $20.50\pm0.08$ 	& $13.27\pm0.04$ 	& $-3.06\pm0.09$ 		& $-2.04\pm0.09$ 		& $-1.02\pm0.05$ 	& O  & $-2.13\pm0.09$   & $-0.93\pm0.06$ 	& 7 \\
Q\,1055+4611 	& 3.317 	& $20.34\pm0.10$ 	& $<14.09$ 		& $<-2.08$ 			& $>-1.56$ 		 	& $<-0.52$ 	 	& Si &$>-1.56$ 	   	& $<-0.52$ 	 	& 33 \\
B\,1104-181 	& 1.661 	& $20.85\pm0.01$ 	& $15.04\pm0.16$ 	& $-1.64\pm0.16$ 		& $-1.06\pm0.02$ 		& $-0.58\pm0.16$ 	& O  & $-0.74\pm0.20$   & $-0.90\pm0.26$ 	& 34 \\
Q\,1108-077	& 3.608 	& $20.37\pm0.07$ 	& $<13.22$ 		& $<-2.98$ 			& $-1.53\pm0.07$ 		& $<-1.39$ 	 	& O  & $-1.69\pm0.08$   & $<-1.20$ 	 	& 18 \\
J\,1111+1332 	& 2.271 	& $20.39\pm0.04$ 	& $13.53\pm0.02$ 	& $-2.69\pm0.04$ 		& $-1.95\pm0.04$ 		& $-0.74\pm0.03$ 	& O  & $-1.92\pm0.08$   & $-0.77\pm0.07$ 	&  7 \\
{\bf J\,1113-1533} & 3.265 	& $21.30\pm0.05$  	& $<14.23$ 		& $<-2.90$ 			& $<-1.53$ 			& $<-1.37$ 	 	& S  & $-1.75\pm0.07$   & $<-1.00$ 	 	& 4, 28 \\
BR\,1117-1329 	& 3.351 	& $20.84\pm0.12$ 	& $<14.53$ 		& $<-2.14$ 			& $-1.22\pm0.13$ 	 	& $<-0.77$ 	 	& Si & $-1.22\pm0.13$   & $<-0.77$ 	 	& 35 \\
B\,1122-168 	& 0.681		& $20.45\pm0.15$ 	& $<14.50$ 		& $<-1.78$ 			& $-0.60\pm0.19$ 	 	& $<-0.82$ 	 	& Si & $-0.60\pm0.19$   & $<-0.82$ 	 	& 36 \\
HS\,1132+2243 	& 2.783 	& $21.00\pm0.07$ 	& $14.02\pm0.02$ 	& $-2.81\pm0.07$ 		& $-2.02\pm0.14$ 		& $-0.79\pm0.12$ 	& S  & $-2.05\pm0.09$   & $-0.76\pm0.06$ 	& 12 \\
\hline
\end{tabular}			       			 	 
\end{center}			       			 	 
\end{minipage}
\end{table*}

\addtocounter{table}{-1}
\begin{table*}
\begin{minipage}{180mm}
\begin{center}
\setlength{\tabcolsep}{2.6pt}
\caption{Continued.}
\begin{tabular}{l c c c c c c c c c c}
\hline\hline                       
QSO & $z_{abs}$ &  log $N$(\hi) & log $N$(\ni) & [N/H]  & [Si/H]	& [N/Si] & $\alpha$  & [$\alpha$/H] & [N/$\alpha$] & Ref. \\
	 & & cm$^{-2}$ 	& cm$^{-2}$ &  \\ 
\hline
{\bf J\,1155+0530} & 2.608 	& $20.37\pm0.11$ 	& $13.76\pm0.06$ 	& $-2.44\pm0.13$ 		& $-1.57\pm0.12$ 		& $-0.87\pm0.10$ 	& S  & $-1.75\pm0.12$   & $-0.69\pm0.09$ 	& 4 \\	
B\,1202-074 	& 4.384 	& $20.49\pm0.16$ 	& $13.80\pm0.10$ 	& $-2.52\pm0.19$ 		& $-1.59\pm0.17$ 	 	& $-0.93\pm0.12$ 	& Si & $-1.59\pm0.17$   & $-0.93\pm0.12$ 	& 37 \\
LBQS\,1210+1731 	& 1.892 	& $20.70\pm0.08$ 	& $14.71\pm0.09$ 	& $-1.82\pm0.12$ 		& $-0.93\pm0.08$ 		& $-0.90\pm0.09$ 	& S  & $-0.86\pm0.09$   & $-0.96\pm0.09$ 	& 18, 21 \\
J\,1211+0422 	& 2.377 	& $20.80\pm0.10$ 	& $14.30\pm0.05$ 	& $-2.33\pm0.11$ 		& $-1.40\pm0.11$ 		& $-0.93\pm0.06$ 	& S  & $-1.39\pm0.11$   & $-0.94\pm0.06$ 	& 38 \\
{\bf B\,1220-1800} & 2.113 	& $20.12\pm0.07$ 	& $13.93\pm0.06$ 	& $-2.02\pm0.09$ 		& $>-0.88$ 			& $<-1.32$ 	 	& S  & $-0.87\pm0.08$   & $-1.15\pm0.07$ 	& 4 \\
LBQS\,1223+1753 	& 2.466 	& $21.40\pm0.10$ 	& $14.83\pm0.17$ 	& $-2.40\pm0.20$ 		& $-1.44\pm0.10$ 		& $-0.96\pm0.17$ 	& S  & $-1.38\pm0.11$   & $-1.02\pm0.17$ 	& 5, 18 \\
LBQS\,1232+0815 	& 2.334 	& $20.90\pm0.08$ 	& $14.63\pm0.08$ 	& $-2.10\pm0.11$ 		& $-1.17\pm0.14$ 		& $-0.93\pm0.14$ 	& S  & $-1.19\pm0.13$   & $-0.91\pm0.13$ 	& 22, 39 \\
{\bf LBQS\,1242+0006} & 1.824 	& $20.45\pm0.10$  	& $>14.20$ 		& $>-2.08$ 			& $-1.20\pm0.11$ 		& $>-1.00$ 	 	& S  & $-1.09\pm0.11$   & $>-1.11$ 	 	& 4 \\
B\,1331+170 	& 1.776 	& $21.15\pm0.07$ 	& $<15.23$ 		& $<-1.75$ 			& $-1.38\pm0.07$ 		& $<-0.35$ 	 	& S  & $-1.19\pm0.13$   & $<-0.23$ 	 	& 8, 9 \\
J\,1337+3152 	& 3.174 	& $21.36\pm0.10$ 	& $14.82\pm0.06$ 	& $-2.36\pm0.12$ 		& $-1.37\pm0.11$ 		& $-0.99\pm0.08$ 	& O  & $-1.29\pm0.12$   & $-1.07\pm0.08$ 	& 40 \\
J\,1337+3152 	& 3.168 	& $20.41\pm0.15$ 	& $<12.80$ 		& $<-3.44$ 			& $-2.68\pm0.16$ 		& $<-0.61$ 	 	& O  & $-2.67\pm0.17$   & $<-0.50$ 	 	& 40 \\
J\,1340+1106 	& 2.508 	& $20.09\pm0.05$ 	& $12.80\pm0.04$ 	& $-3.12\pm0.06$ 		& $-1.85\pm0.05$ 		& $-1.27\pm0.04$ 	& O  & $-1.76\pm0.06$   & $-1.36\pm0.05$ 	& 26 \\
J\,1340+1106 	& 2.796 	& $21.00\pm0.06$ 	& $14.04\pm0.02$ 	& $-2.79\pm0.06$ 		& $-1.83\pm0.06$ 		& $-0.96\pm0.03$ 	& O  & $-1.65\pm0.07$   & $-1.14\pm0.04$ 	& 26 \\
J\,1342-1355 	& 3.118 	& $20.05\pm0.08$ 	& $13.28\pm0.02$ 	& $-2.60\pm0.08$ 		& $-1.15\pm0.09$ 		& $-1.45\pm0.04$ 	& O  & $-1.22\pm0.08$   & $-1.38\pm0.03$ 	& 11 \\
B\,1409+0930 	& 2.456 	& $20.53\pm0.08$ 	& $<13.19$ 		& $<-3.17$ 			& $-1.96\pm0.08$ 		& $<-1.15$ 	 	& O  & $-2.07\pm0.10$   & $<-0.92$ 	 	& 41 \\
B\,1409+0930 	& 2.668 	& $19.80\pm0.08$ 	& $<13.49$ 		& $<-2.14$ 			& $-1.18\pm0.09$ 		& $<-0.87$ 	 	& O  & $-1.18\pm0.14$   & $<-0.63$ 	 	& 9, 41 \\
J\,1419+0829 	& 3.050 	& $20.40\pm0.03$ 	& $13.28\pm0.02$ 	& $-2.95\pm0.04$ 		& $-2.08\pm0.03$ 		& $-0.87\pm0.02$ 	& O  & $-1.93\pm0.03$   & $-1.02\pm0.02$ 	& 7, 30 \\
Q\,1425+6039 	& 2.826	& $20.30\pm0.04$ 	& $14.71\pm0.01$ 	& $-1.42\pm0.04$ 		& $>-0.98$ 		 	& $<-0.47$ 	 	& S & $-1.26\pm0.12$ 	&  $-0.16\pm0.10$ 	 & 25 \\	
J\,1435+5359 	& 2.343 	& $21.05\pm0.10$ 	& $14.78\pm0.05$ 	& $-2.21\pm0.10$ 		& $-1.43\pm0.10$ 		& $-0.78\pm0.03$ 	& S  & $-1.39\pm0.11$   & $-0.82\pm0.05$ 	& 28 \\
J\,1439+1117 	& 2.418 	& $20.10\pm0.10$ 	& $>15.71$ 	& $>-0.22$ 		& $-0.81\pm0.11$ 			& $>0.47$ 	 	& S  & $0.05\pm0.12$   & $>-0.45$ 	& 42 \\
J\,1443+2724 	& 4.224 	& $20.95\pm0.10$ 	& $15.52\pm0.01$ 	& $	-1.26\pm0.10$ 		& $>-1.03$ 			& $<-0.26$ 	 	& S  & $-0.55\pm0.10$   & $-0.71\pm0.01$ 	& 10, 18 \\
J\,1558+4053 	& 2.553 	& $20.30\pm0.04$ 	& $12.66\pm0.07$ 	& $-3.47\pm0.08$ 		& $-2.49\pm0.04$ 		& $-0.98\pm0.07$ 	& O  & $-2.45\pm0.06$   & $-1.02\pm0.08$ 	& 26 \\
J\,1558-0031 	& 2.703 	& $20.67\pm0.05$ 	& $14.45\pm0.02$ 	& $-2.05\pm0.05$ 		& $-1.69\pm0.06$ 		& $-0.36\pm0.04$ 	& O  & $-1.57\pm0.06$   & $-0.48\pm0.04$ 	& 30, 43 \\
GB\,1759+7539 & 2.625 	& $20.80\pm0.01$ 	& $14.99\pm0.03$ 	& $-1.64\pm0.03$ 		& $-0.79\pm0.01$ 		& $-0.85\pm0.03$ 	& S  & $-0.71\pm0.02$   & $-0.93\pm0.03$ 	& 36, 44 \\
Q\,1946+7658 	& 2.844 	& $20.27\pm0.06$ 	& $12.59\pm0.04$ 	& $-3.51\pm0.07$ 		& $-2.18\pm0.06$ 		& $-1.36\pm0.04$ 	& O  & $-2.14\pm0.06$   & $-1.37\pm0.04$ 	& 2, 18 \\
Q\,2059-360 	& 2.507  	& $20.29\pm0.07$ 	& $12.66\pm0.18$ 	& $-3.46\pm0.19$ 		& $-2.07\pm0.13$ 	 	& $-1.39\pm0.21$ 	& S & $-1.77\pm0.20$   & $-1.69\pm0.26$ 	& 27, 45 \\
Q\,2059-360 	& 3.083 	& $20.98\pm0.08$ 	& $13.95\pm0.02$ 	& $-2.86\pm0.08$ 		& $-1.63\pm0.09$ 		& $-1.23\pm0.04$ 	& O  & $-1.58\pm0.09$   & $-1.28\pm0.04$ 	& 11 \\
{\bf J\,2119-3536 }& 1.996 	& $20.10\pm0.07$  	& $<14.70$ 		& $<-1.23$ 		& $-0.28\pm0.08$ 		& $<-1.07$ 	 	& Si  & $-0.28\pm0.08$ 		& $<-1.07$	    	& 4, 9 \\	
{\bf LBQS\,2138-4427} & 2.852 	& $20.98\pm0.05$ 	& $14.27\pm0.04$ 	& $-2.54\pm0.06$ 		& $-1.63\pm0.05$ 		& $-0.91\pm0.04$ 	& S  & $-1.60\pm0.05$   & $-0.94\pm0.04$ 	& 4, 5 \\
Q\,2206-1958A 	& 2.076 	& $20.43\pm0.05$ 	& $12.79\pm0.05$ 	& $-3.47\pm0.07$ 		& $-2.29\pm0.05$ 		& $-1.18\pm0.05$ 	& O  & $-2.07\pm0.06$   & $-1.40\pm0.06$ 	& 29 \\
Q\,2212-1626	& 3.662		& $20.20\pm0.08$ 	& $<13.58$ 		& $<-2.45$ 			& $-1.80\pm0.08$ 	 	& $<-0.62$ 	 	& Si & $-1.80\pm0.08$   & $<-0.62$ 	 	& 23, 33 \\
{\bf B\,2222-396} 	& 2.154 	& $20.85\pm0.10$  	& $14.34\pm0.05$ 	& $-2.34\pm0.11$ 		& $>-1.32$ 			& $<-1.17$ 	 	& S  & $-1.72\pm0.12$   & $-0.62\pm0.08$ 	& 4, 10 \\
Q\,2223+20 	& 3.119 	& $20.30\pm0.10$ 	& $<13.62$ 		& $<-2.51$ 			& $-2.17\pm0.11$ 		& $<-0.23$ 	 	& Si  &  $-2.17\pm0.11$ 		& $<-0.23$ 	    	& 31 \\
LBQS\,2230+0232 	& 1.864 	& $20.90\pm0.10$ 	& $14.90\pm0.06$ 	& $-1.83\pm0.12$ 	& $-0.75\pm0.10$ 		&$-1.08\pm0.06$  	& S  & $-0.73\pm0.14$   & $-1.10\pm0.12$ 	& 8, 21 \\
Q\,2231-002 	& 2.066 	& $20.56\pm0.10$ 	& $<15.02$ 		& $<-1.37$ 			& $-0.83\pm0.10$ 		& $<-0.48$ 	 	& S  & $-0.58\pm0.21$   & $<-0.25$ 	 	& 28 \\
Q\,2233+1310 	& 3.149 	& $20.00\pm0.10$ 	& $<14.32$ 		& $<-1.51$ 			& $>-1.00$ 		 	& $<-0.51$ 	 	& Si & $>-1.00$ 	& $<-0.51$ 	    	& 33 \\
J\,2241+1352 & 4.282	& $21.15\pm0.10$ & $>14.81$ & $>-2.17$ &	$>-1.65$ & 	$>-0.52$ & 	S & $-1.69\pm0.10$ & 	$>-0.57$ & 12\\
HE\,2243-6031 	& 2.330	& $20.67\pm0.02$ 	& $14.88\pm0.10$ 	& $-1.62\pm0.10$ 		& $-0.82\pm0.03$ 		& $-0.80\pm0.10$ 	& S  & $-0.77\pm0.03$   & $-0.85\pm0.10$ 	& 46 \\	
B\,2318-1107 	& 1.989 	& $20.68\pm0.05$ 	& $>14.55$ 		& $>-1.96$ 			& $-0.85\pm0.05$ 		& $>-1.14$ 	 	& S  & $-0.71\pm0.06$   & $>-1.37$ 	 	& 47 \\
B\,2332-094 	& 3.057	& $20.50\pm0.07$ 	& $13.73\pm0.03$ 	& $-2.60\pm0.08$ 		& $\cdots$ 			& $\cdots$ 	 	& O  & $-1.24\pm0.07$   & $-1.36\pm0.04$ 	& 11 \\
B\,2342+3417 	& 2.908 	& $21.10\pm0.10$ 	& $14.92\pm0.04$ 	& $-2.01\pm0.11$ 		& $-0.99\pm0.10$ 		& $-1.02\pm0.04$ 	& S  & $-1.03\pm0.10$  & $-0.98\pm0.04$ 	& 25 \\
B\,2343+125 	& 2.431 	& $20.40\pm0.07$ 	& $14.62\pm0.01$ 	& $-1.61\pm0.07$ 		& $-0.76\pm0.08$ 		& $-0.85\pm0.03$ 	& O  & $-0.64\pm0.09$   & $-0.97\pm0.05$ 	& 22, 48 \\
J\,2344+0342	& 3.219	& $21.35\pm0.07$	& $>14.65$ & $>-2.53$ & 	$>-1.99$	& 	$>-0.54$ & 	Si & $>-1.99$	& 	$>-0.54$ &  12 \\
Q\,2344+1228 	& 2.538	& $20.36\pm0.10$ 	& $13.78\pm0.03$ 	& $-2.41\pm0.10$ 		& $-1.65\pm0.10$ 	 	& $-0.76\pm0.03$ 	& Si & $-1.65\pm0.10$   & $-0.76\pm0.03$     	& 2 \\
J\,2346+1247 	& 2.569 	& $20.98\pm0.04$ 	& $15.02\pm0.04	$ 	& $-1.79\pm0.06$ 		& $-0.70\pm0.08$ 		& $-1.09\pm0.08$ 	& S  & $-0.72\pm0.06$   & $-1.07\pm0.06$ 	& 49 \\
{\bf B\,2348-0180} & 2.615 	& $21.30\pm0.08$  	& $14.46\pm0.05$ 	& $-2.67\pm0.09$ 		& $-1.81\pm0.10$ 	 	& $-0.86\pm0.08$ 	& Si & $-1.81\pm0.10$   & $-0.86\pm0.08$ 	& 4, 18 \\
B\,2348-147 	& 2.279 	& $20.56\pm0.08$ 	& $<13.22$ 		& $<-3.17$  			& $-1.87\pm0.08$ 		& $<-1.27$ 	 	& S  & $-1.96\pm0.14$   & $<-0.86$ 	 	& 8, 50 \\
{\bf LBQS\,2359-0216} & 2.154 	& $20.30\pm0.10$ 	& $<14.10$ 		& $<-2.03$ 			& $-1.46\pm0.10$ 	 	& $<-0.48$ 	 	& Si & $-1.46\pm0.10$   & $<-0.48$ 	 	& 4, 8 \\
\hline				       			 	 
\end{tabular}			       			 	 
\end{center}			       			 	 
\end{minipage}
\begin{minipage}{180mm}
{\bf References:} 
1: \citet{molaro01};
2: \citet{prochaska02};
3: \citet{richter05}; 
4: This work; 
5: \citet{srianand05}; 
6: \citet{noterdaeme07}; 
7: \citet{cooke14}; 
8: \citet{prochaska98}; 
9: \citet{dessauges03}; 
10: \citet{noterdaeme08}; 
11: \citet{petitjean08}; 
12: \citet{prochaska03}; 
13: \cite{ellison10};
14: \citet{zafar11}; 
15: \citet{ellison01}; 
16: \citet{dutta14};
17: \citet{dessauges01};
18: \citet{prochaska01}; 
19: \citet{ledoux03}; 
20: \citet{lopez03}; 
21: \citet{dessauges06}; 
22: \citet{centurion03}; 
23: \citet{lu96}; 
24: \citet{srianand98}; 
25: \citet{prochaska07}; 
26: \citet{cooke11}; 
27: \citet{ledoux06}; 
28: \citet{henry07}; 
29: \citet{pettini08}; 
30: Cooke (Pvt. comm.); 
31: \citet{prochaska06}; 
32: \citet{meiring11}; 
33: \citet{lu98};
34: \citet{lopez99}; 
35: \citet{peroux02};
36: \citet{delaverga00};
37: \citet{lu96a}; 
38: \citet{lehner08};
39: \citet{srianand00}; 
40: \citet{srianand10}; 
41: \citet{pettini02}; 
42: \citet{noterdaeme08b};
43: \citet{omeara06};
44: \citet{outram99};
45: \citet{dessauges03b};
46: \citet{lopez02};
47: \citet{noterdaeme07}; 
48: \citet{dodorico02};
49: \citet{rix07}; 
50: \citet{dessauges07} \\
$^{\dagger}$ The column densities are provided for the component seen in \ni.
\end{minipage}
\end{table*}			       			 	 


\bibliographystyle{aa}
\bibliography{nitrogen.bib}{}

\bsp

\label{lastpage}
\end{document}